\documentclass[smallextended]{svjour3}     % onecolumn (second format)
\usepackage{colortbl,ulem,arydshln,natbib}
\usepackage{dsfont}
\usepackage{longtable}
\usepackage[hang,small,bf]{caption}
\usepackage{multirow}
\usepackage{graphicx}
\usepackage{epstopdf}
\usepackage{textcomp}
\usepackage{pdfpages}
\usepackage[utf8]{inputenc}
\usepackage[english]{babel}
\usepackage{amsmath,amssymb,latexsym,float}
\usepackage{array}
\usepackage{graphicx,color}
\DeclareGraphicsExtensions{.pdf,.eps,.png,.jpeg}
\newcommand{\bfX}{{\bf X}}
\newcommand{\bfD}{{\bf D}}
\newcommand{\bfU}{{\bf U}}

\newcommand{\bfM}{{\bf M}}

\newcommand{\bfW}{{\bf W}}
\newcommand{\bfV}{{\bf V}}
\newcommand{\bfx}{{\bf x}}
\newcommand{\bfz}{{\bf z}}

\newcommand{\bff}{{\bf f}} 

\journalname{Statistical Theory and Methods}
\begin{document}
\title{A principal components method to impute missing values for mixed data}

\author{Vincent Audigier \and Fran\c cois Husson \and Julie Josse}

\institute{
              Agrocampus Ouest, 65 rue de St-Brieuc, F-35042 Rennes \\
              Tel.: +33-223485874\\
              Fax: +33-223485871\\
              \email{husson@agrocampus-ouest.fr}           %  \\
}

\date{Received: date / Accepted: date}

\maketitle

\begin{abstract}
We propose a new method to impute missing values in mixed datasets. It is based on a principal components method, the factorial analysis for mixed data, which balances the influence of all the variables that are continuous and categorical in the construction of the dimensions of variability. Because the imputation uses the principal axes and components, the prediction of the missing values is based on the similarity between individuals and on the relationships between variables. 
The properties of the method are illustrated via simulations and the quality of the imputation is assessed through real datasets. The method is compared to a recent method \citep{Buhlmann11} based on random forests and shows better performances especially for the imputation of categorical variables and when there are highly linear relationships between continuous variables.
\end{abstract}

\section{Introduction}
Missing data are a key problem in statistical practice. Indeed, they are never welcome, because most statistical methods cannot be applied directly on an incomplete dataset. One of the common approaches to deal with missing values consists in imputing missing values by plausible values. It leads to a complete dataset that can be analysed by any statistical method. However, the interpretation of the results should be done with caution, since there is necessarily uncertainty associated with the prediction of values.

Several imputation methods are available for continuous data such as the K-nearest neighbours imputation \citep{Troyanskaya01}, the imputation based on the multivariate normal model \citep{Schafer97} or the multivariate imputation by chained equations \citep{Buuren99,Buuren07}. The multivariate normal model supposes to define a joint distribution for the data which in practice can be restrictive. 
The imputation by chained equation consists in defining a model for each variable with missing data that can afford a finer modelling but requires defining many models. It is also possible to impute continuous data with principal component analysis (PCA). The idea is to use an algorithm that performs PCA despite the missingness of some data \citep{Kiers97,Josse09,Ilin10}. Principal components and axes obtained by this algorithm are then used to reconstruct the data, which provides an imputation of the missing entries. This method has the particular advantage of simultaneously imputing any missing data by taking into account the similarities between individuals and the relationships between variables. There was great excitement around this approach especially in the machine learning community in matrix completion problems such as the Netflix challenge \citep{Netflix}.

The imputation of categorical data can also be done with non-parametric methods such as the K-nearest neighbours or with parametric methods thanks to different models. The most classical model is the log-linear one \citep{Schafer97}. It has a major drawback: the number of parameters increases rapidly with the number of levels of categorical variables. Therefore, in practice, it becomes unusable in some cases. Other models have been proposed to overcome this problem, such as the latent class model \citep{Vermunt08}, or the log-linear hierarchical model \citep{Schafer97,Little02}.

Finally, for mixed data, \textit{i.e.} both continuous and categorical, literature is less abundant. A solution consists in coding the categorical variables using the indicator matrix of dummy variables, and using an imputation method dedicated to continuous variables on the concatenated matrix of continuous variables and the indicator matrix. However, this method is not satisfactory since the hypotheses on continuous variables cannot be assumed for dummy variables. \citet{Schafer97} proposed an imputation based on the general location model which can be seen as combination of log linear model and multivariate normal model with the benefits and drawbacks of such models. 
The imputation by chained equations \citep{Buuren99,Buuren07} is one of the only approaches that can be easily extended to the mixed case by defining a specific model for each continuous and each categorical variables. However, as mentioned for the continuous case, many models have to be defined. 
Recently, \citet{Buhlmann11} proposed an imputation method based on random forests \citep{Breiman01}. The imputation is done using the following iterative algorithm: after replacing the missing data with initial values, missing values of the variable with the least number of missing values are predicted by random forests. This operation is performed for each variable in the dataset and the procedure is repeated until the predictions stabilize. For mixed data, they compared their method to the imputation by chained equation and to a version of the K-nearest neighbours method adapted for mixed data. Their approach clearly outperforms the competing methods regarding many simulations and real datasets. It provides a good quality of imputation whatever the number of individuals and of variables, regardless of the type of relationship between variables, and it is little sensitive to the tuning parameters. That is why this nonparametric method based on random forests can serve as a reference among the existing methods.

We propose a new imputation method for mixed data based on a principal components method dedicated to mixed data: the factorial analysis for mixed data (FAMD) presented in \citet{Escofier79} and also named PCAMIX \citep{Kiers91}. 
We begin by presenting the imputation method (section 2) and then focus on its properties that are illustrated by simulations (section 3). Finally, the method is assessed on real datasets (section 4). The competitiveness of the method is highlighted by comparing its performances to the ones of \citet{Buhlmann11} method.
\section{Imputation for mixed type-data thanks to FAMD}
\subsection{FAMD in the complete case \label{FAMD_complete_case}}
FAMD is a principal components method to describe, summarise and visualise multidimensional matrix with mixed data. As any principal components method, its aim is to study the similarities between individuals, the relationships between variables (here continuous and categorical variables) and to link the study of the individuals with that of the variables. Such methods reduce the dimensionality of the data and provide subspace that best represent the data. More precisely, the subspace is obtained maximising the variability of the projected points. Dimensionality reduction is achieved through the singular value decomposition (SVD) of specific matrices. As mentioned by \citet{Benz73}: ``Doing a data analysis, in good mathematics, is simply searching eigenvectors, all the science of it (the art) is just to find the right matrix to diagonalize''.

The principle of FAMD is to balance the influence of the continuous and the categorical variables in the analysis.
The rationale is to weight the variables in such a way that each variable of both types contributes equivalently to the construction of the dimensions of variability. It is the same idea as scaling for continuous variables in PCA.
Let us note $I$ the number of individuals, $K_1$ the number of continuous variables, $K_2$ the number of categorical variables and $K=K_1+K_2$ the total number of variables. The first step of FAMD consists in coding the categorical variables using the indicator matrix of dummy variables. 
We note $\bfX_{I \times J}$ the matrix where $\left(\bfx_j\right)_{1\leq j\leq K_1}$ are continuous variables and $\left(\bfx_j\right)_{K_1+1\leq j \leq J}$ are dummy variables. The total number of columns is $J=K_1+\sum_{k=K_1+1}^{K}{q_k}$ where $q_k$ is the number of categories of the variable $k$ (noted $\bfz_k$). The second step is a weighting step: each continuous variable $\bfx_{j}$ is standardized (divided by its standard deviation $s_j$) and each dummy variable is divided by $\sqrt{p_{j}}$ where $p_{j}$ denotes the proportion of individuals that take the category $j$ ($j=K_1+1,\ldots,J$).
Finally FAMD consists in performing a PCA on the weighted matrix $\bfX\bfD_{\Sigma}^{-1/2}$, where $\bfD_{\Sigma}$ is the diagonal matrix $diag\left(s_{x_1}^2,...,s_{x_{K_1}}^2,p_{K_1+1},\ldots,p_j,\ldots,p_{J}\right)$. It boils down to performing the singular value decomposition of the matrix $\left(\bfX\bfD_{\Sigma}^{-1/2}-\bfM\right)$ with $\bfM_{I\times J}$ the matrix with each row equals to the vector of the means of each column of $\bfX\bfD_{\Sigma}^{-1/2}$. As in any principal components methods, the first $S$ dimensions of variability are preserved.

The specific weighting implies that the distances between two individuals $i$ and $i'$ in the initial space (before approximating the distances by keeping the first $S$ dimensions obtained from the SVD) is as follow:
\begin{eqnarray*}
d^2\left(i,i'\right)&=&\sum_{k=1}^{K_1}\frac{\left(x_{ik}-x_{i'k}\right)^2}{s_{\bfx_k}^2}+\sum_{j=K_1+1}^J\frac{1}{p_{j}}\left(x_{ij}-x_{i'j}\right)^2
\end{eqnarray*}
Weighting by $\frac{1}{s_{\bfx_k}^2}$ ensures that units of continuous variables don't influence the (square) distance between individuals. Furthermore, weighting by $\frac{1}{p_{j}}$ implies that two individuals taking different categories for the same variable are more distant when one of them takes a rare category than when both of them take frequent categories. The margins of the categorical variables play an important role in this method.
It can be seen regarding the variance in the initial space, also called inertia, of the category $j$ \citep{Escofier79}: $\mbox{Inertia}(\bfx_j)=1-p_j$. Categories with a small frequency have a greater inertia than the others and consequently rare categories have a greater influence on the construction of the dimensions of variability.

The specific weighting implies also that, in FAMD, the principal components maximise the links with both continuous and categorical variables. More precisely, the first principal component $\bff_1$ maximises
\begin{eqnarray}\label{crit.r2.eta2}
\sum_{k=1}^{K_1}R^2\left(\bfx_k,\bff_1\right)+\sum_{k=K_1+1}^K\eta^2\left(\bfz_k,\bff_1\right)
\end{eqnarray}
It is the synthetic variable the most linked with, on the one hand the continuous variables within the meaning of the coefficient of determination ($R^2$), and on the other hand, with the categorical variables within the meaning of the squared correlation ratio ($\eta^2$). The second principal component is the synthetic variable which maximises the criterion among orthogonal variables to the first principal component, etc. 

Regarding the criterion~\eqref{crit.r2.eta2}, we can note that when there are only continuous variables, FAMD reduces to PCA whereas when there are only categorical variables, FAMD reduces to multiple correspondence analysis \citep{Leb84,Green06}.

\subsection{The iterative FAMD algorithm}
There exists a method to impute continuous data based on PCA \citep{Kiers97,Josse09}.\label{algo_FAMD}
The algorithm boils down to setting the missing elements at initial values, performing the PCA on the completed dataset, imputing the missing values with values predicted by the reconstruction formula (defined by the fitted matrix obtained with the axes and components) using a predefined number of dimensions, and repeating the procedure on the newly obtained matrix until the total change in the matrix falls below an empirically determined threshold. 

Since FAMD has been presented as the PCA of the matrix $\left(\bfX\bfD_{\Sigma}^{-1/2}\right)$, the methodology proposed to impute data with PCA can be extended to FAMD, but it is necessary to adapt the algorithm to take into account the specificities of FAMD. More precisely, the same algorithm can be used but the matrix $\bfD_{\Sigma}$ as well as the mean matrix $\bfM$ have to be updated during the estimation process because they depend on all the data. Indeed, after imputing data with the reconstruction formula, the variance of the continuous variables as well as the column margins of the categorical variables of the new data table have changed. 

\begin{figure}[!hbt]
\begin{center}
\includegraphics[width=0.9\textwidth]{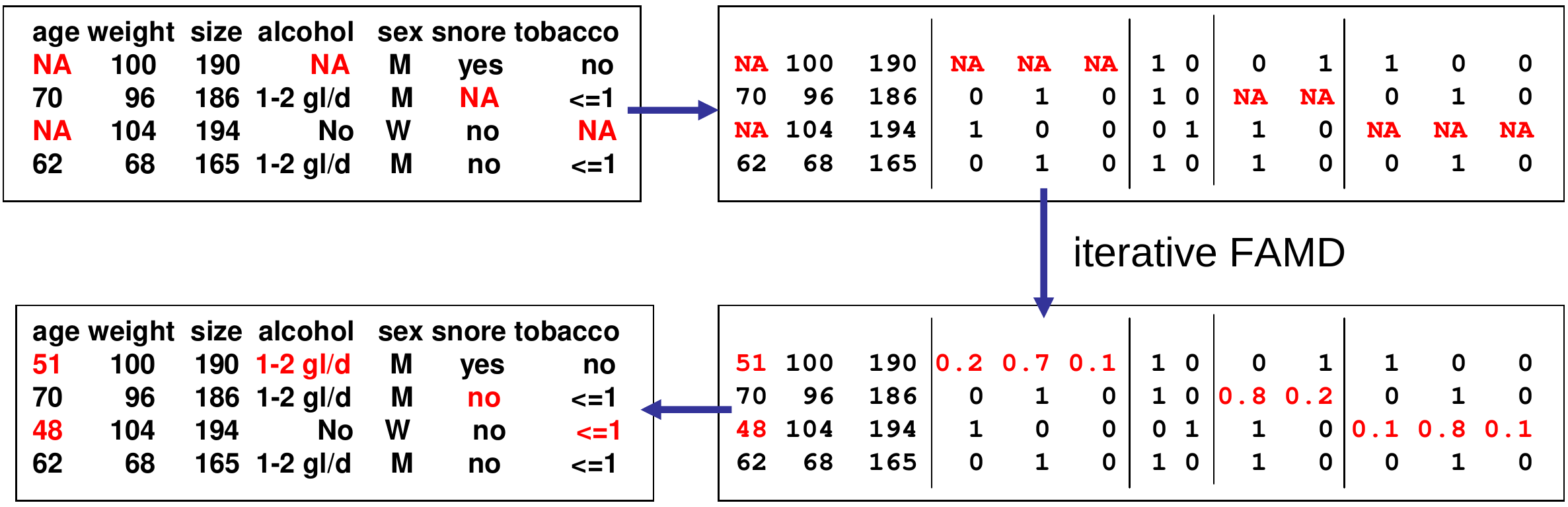}
\caption{Diagram for the iterative FAMD algorithm: the raw mixed data, the matrix $\bfX$, the imputed data obtained by iterative FAMD and the imputed mixed data.\label{diagram}}
\end{center}
\end{figure}

The algorithm to impute missing values for mixed data based on FAMD starts as follow. There is initially a table of mixed data with missing values (first table in Figure~\ref{diagram}). This table is then transformed to obtain the matrix $\bfX$ coding categorical variables using an indicator matrix of dummy variables. A missing value on a categorical variable then leads to a row of missing values in the indicator matrix (second table in Figure~\ref{diagram}). Then this data table is imputed according to the following algorithm. We denote by $\bfW_{I\times J}$ the matrix of weights such that $w_{ij}$ is equal to 1 if $x_{ij}$ is observed and 0 otherwise. The Hadamard product is denoted by $*$ and ${\bf 1}_{I\times J}$ is a matrix with only ones. The imputation algorithm based on FAMD called iterative FAMD, is then written as follows:

\begin{enumerate}
\item initialization $\ell=0$: substitute missing values by initial values as, for example, the mean of the variable for the continuous variables and the proportion of the category for each category using the non-missing entries. Note that concerning the categorical variables, the initial values can be noninteger ones but the sum of the entries corresponding to one individual and one categorical variable must equal one.\\
Calculate $\bfD_{\Sigma}^0$ and $\bfM^0$ the mean of $\bfX^0\left(\bfD_{\Sigma}^0\right)^{-1/2}$.
\item step $\ell$:
\begin{itemize}
\item[(a)] perform the FAMD which boils down to performing the singular value decomposition of $\left(\bfX^{\ell -1}\left(\bfD_{\Sigma}^{\ell -1}\right)^{-1/2}-\bfM^{\ell-1}\right)$ to obtain the matrices $\hat \bfU^{\ell}$ and $\hat \bfV^{\ell}$ with the left and right singular vectors as well as the matrix $\left(\hat{\bf\Lambda}^{\ell}\right)^{1/2}$
with the singular values;
\item[(b)] keep the first $S$ dimensions and use the reconstruction formula to compute the fitted matrix: 
$$\hat{\bfX}^{\ell}_{I\times J} = \left(\hat{\bfU}_{I\times S}^{\ell} \left(\hat{\bf\Lambda}_{S\times S}^{\ell}\right)^{1/2}\left(\hat{\bfV}_{J\times S}^{\ell}\right)^{\top}+\bfM^{\ell -1}_{I\times J}\right)\left(\left(\bfD_{\Sigma}^{ \ell -1}\right)_{I\times J}\right)^{1/2}$$
and the new imputed dataset becomes $\bfX^{\ell}= \bfW *\bfX + (\bf1-\bfW)* \hat \bfX^{\ell}$. The observed values are the same but the missing ones are replaced by the fitted values;
\item[(c)] from the new completed matrix $\bfX^{\ell}$, $\bfD_{\Sigma}^{\ell}$ and $\bfM^{\ell}$ are updated.
\end{itemize}
\item steps (2.a), (2.b) and (2.c) are repeated until the change in the imputed matrix falls above a predefined threshold $\sum_{ij}(\hat x_{ij}^{\ell-1}-\hat x_{ij}^{\ell})^2\leq \varepsilon$, with $\varepsilon$ equals to $10^{-6}$ for example.
\end{enumerate}

At the end of the algorithm, imputed values for the missing entries for the categories are not equal to 0 and 1 but are real numbers (third table in Figure~\ref{diagram}). However, the constraint in the initialisation step ensures that the sum of the entries for one individual and one categorical variable is equal to 1. This property comes from the specific weighting and is demonstrated in the framework of multiple correspondence analysis \citep{Tenenhaus85}. Consequently, the imputed values can be considered as degrees of membership to the associated category and it is possible to impute the categorical variable with the most plausible value (last table in Figure~\ref{diagram}).

Such algorithms which alternate a step of estimation of the parameters via a singular value decomposition and a step of imputation of the missing values are known to suffer from overfitting problems. These problems occur even if these methods reduce the dimensionality of the data. In order to avoid such problems, the iterative singular value decomposition algorithm has been replaced by an iterative thresholded singular value decomposition algorithm in the framework of PCA \citep{JosseHusson12,Mazu09}. We follow the approach proposed by \citet{JosseHusson12} and define a regularized iterative FAMD algorithm by replacing the singular values $\left(\sqrt{\hat{\bf\lambda}_{s}^{\ell}}\right)_{s=1,...,S}$ of step (2.b) by $\left(\frac{\hat{\bf\lambda}_{s}^{\ell}-\hat\sigma^2}{\sqrt{\hat{\bf\lambda}_{s}^{\ell}}}\right)_{s=1,...,S}$ with $\hat\sigma^2=\sum_{s=S+1}^{J-K_2}{\frac{\lambda_s}{J-K_2-S}}$. The rationale is to remove the noise in order to avoid instabilities in the prediction. Implicitly, it is assumed that the first $S$ dimensions are made of information and noise whereas the last ones are only noise, that is why the variance of the noise is estimated by the mean of the last eigenvalues. Singular values are thresholded with a greater amount of shrinkage for the smallest ones, which is acceptable since these smallest singular values are more responsible for instability. When the noise is small, the regularized algorithm is quite similar to the non regularized one. When the noise is very important, the regularized algorithm boils down to imputing continuous variables with the mean of the variables and to imputing the categories with the proportion. 

\section{Properties of the imputation method}
We discuss the main properties of the new imputation method and illustrate those properties on different toy datasets. We compare the imputation results obtained with the method based on FAMD to the ones obtained with the method based on random forests \citep{Buhlmann11} to better highlight some properties.

\subsection*{Simulation process}

These toy datasets differ with respect to the number of continuous variables, the number of categorical variables, the number of categories per categorical variable, the number of individuals per categories, the number of underlying dimensions and the intensity of the relationship between variables through different signal to noise ratios (SNR). More precisely, the toy datasets are almost all simulated according to the following procedure:
\begin{itemize}
\item $S^{\prime}$ independent variables are drawn from a standard Gaussian distribution;
\item each variable $s'$ (for $s'= 1,\ldots, S'$) is replicated $K^{s'}$ times which guarantees to obtain (in expectation) orthogonal groups of correlated variables;
\item Gaussian noise is added with different levels of variance to obtain different signal to noise ratios. A high SNR implies that the variables in each group are very linked, whereas a low one corresponds to a very noisy dataset.
\item categorical variables are obtained by splitting continuous variables in equal count categories.
 \end{itemize}
Then we insert missing values completely at random varying the percentage of missing values (10\%, 20\% and 30\%).
For each set of parameters, 200 simulations are done.

\subsection*{Criteria}

Two criteria are used to assess the quality of the imputation, the proportion of falsely classified (PFC) entries for categorical variables and the normalised root mean squared error (NRMSE) for continuous data:
\begin{eqnarray*}
NRMSE&=&\sqrt{\frac{\sum_{k=1}^{K_1}\sum_{i=1}^Iw_{ik}\left(\frac{x_{ik}-\hat{x}_{ik}}{s_{\bfx_k}}\right)^2}{\sum_{k=1}^{K_1}\sum_{i=1}^Iw_{ik}}}
\end{eqnarray*}
$NRMSE$ makes it possible to work on variables with different variances. Moreover, when $NRMSE$ equals zero, the imputation is perfect, whereas when it is close to 1, the imputation gives results similar to those obtained using the mean imputation.

\subsection{Relationships between continuous and categorical variables}
At first the fundamental property of FAMD is to take into account relationships between continuous and categorical variables. 
Taking into account both types of variables improves the imputation as it is illustrated with a dataset that has two underlying dimensions ($S'=2$). Each dimension is composed of two continuous variables and two categorical variables with four categories.
Missing data are then added at random for the three selected percentages and finally the imputation algorithm is performed according to three strategies:
\begin{enumerate}
\item[(1)] using only continuous variables, which leads to an imputation of continuous variables with only those variables;
\item[(2)] using only categorical variables, which leads to an imputation of categorical variables with only those variables;
\item[(3)] using and imputing variables of both kinds.
\end{enumerate}
\begin{figure}[!hbt]
  \centering \includegraphics[scale=0.38]{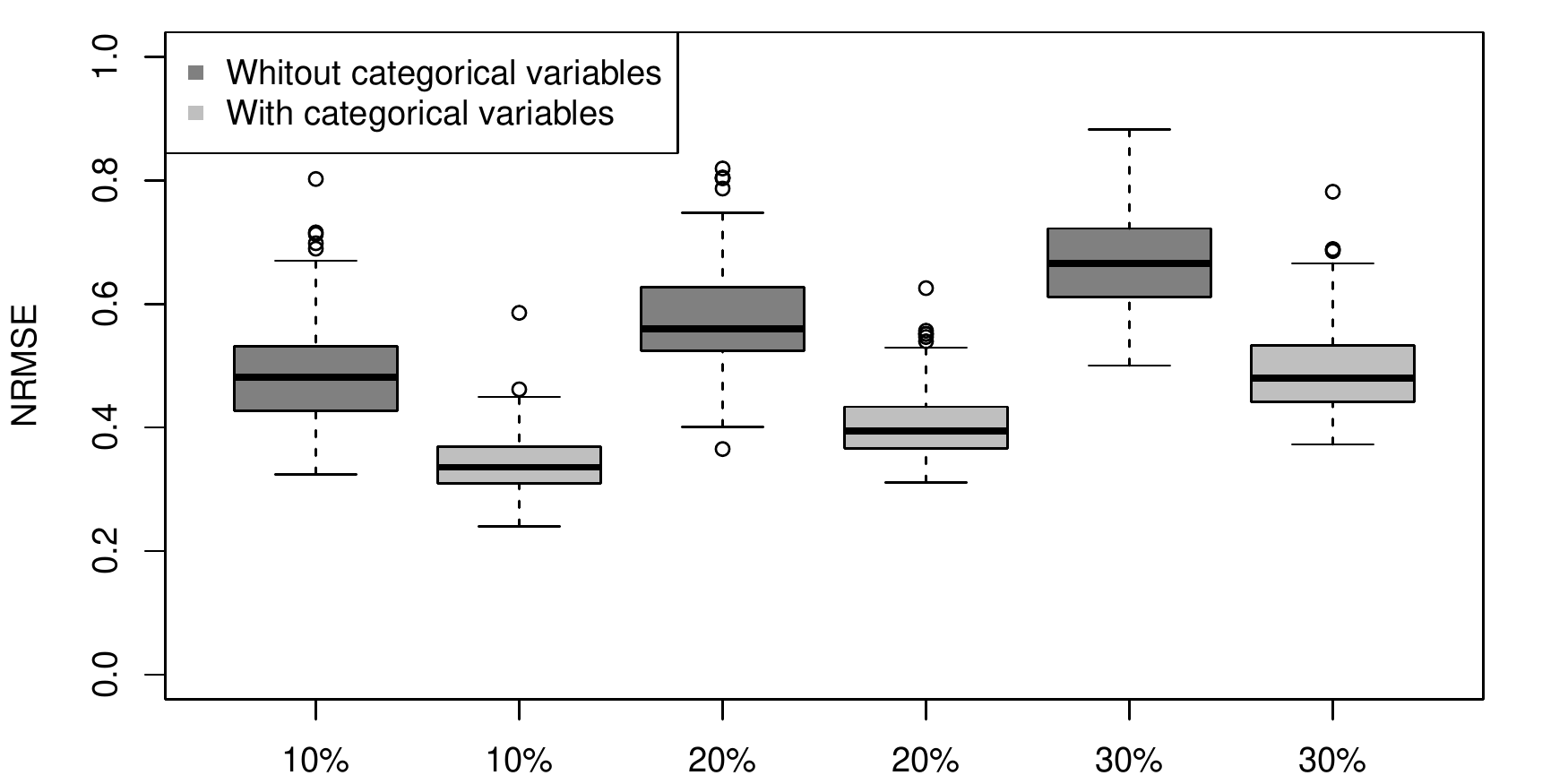}
  \caption{Distribution of the $NRMSE$ for different amounts of missing values (10\%, 20\%, 30\%). Dark grey boxplots correspond to the error of imputation for continuous variables when only continuous variables are used whereas grey boxplots correspond to the error when categorical and continuous variables are used.
  \label{main.prop.1}}
\end{figure}

Figure \ref{main.prop.1} compares the distributions of the NRMSE for the three percentages of missing data according to the strategies (1) and (3). As expected, when the rate of missing data increases, the imputation error is larger. When the SNR decreases, the quality of the imputation (not shown here) decreases which is also expected. It can be noted that even with 30\% of missing data the imputation with iterative FAMD is much better than the mean imputation of the variable (NRMSE less than 1). This is due to the relationships between variables which improve the mean imputation which is the first step of iterative FAMD.
The imputation error is lower when considering both types of variables (boxplots in grey) than when considering only continuous variables (boxplots in dark grey). Taking into account categorical variables thus improves the imputation of continuous variables. This behaviour is the same whatever the rate of missing data.
\begin{figure}[!hbt]
  \centering \includegraphics[scale=0.38]{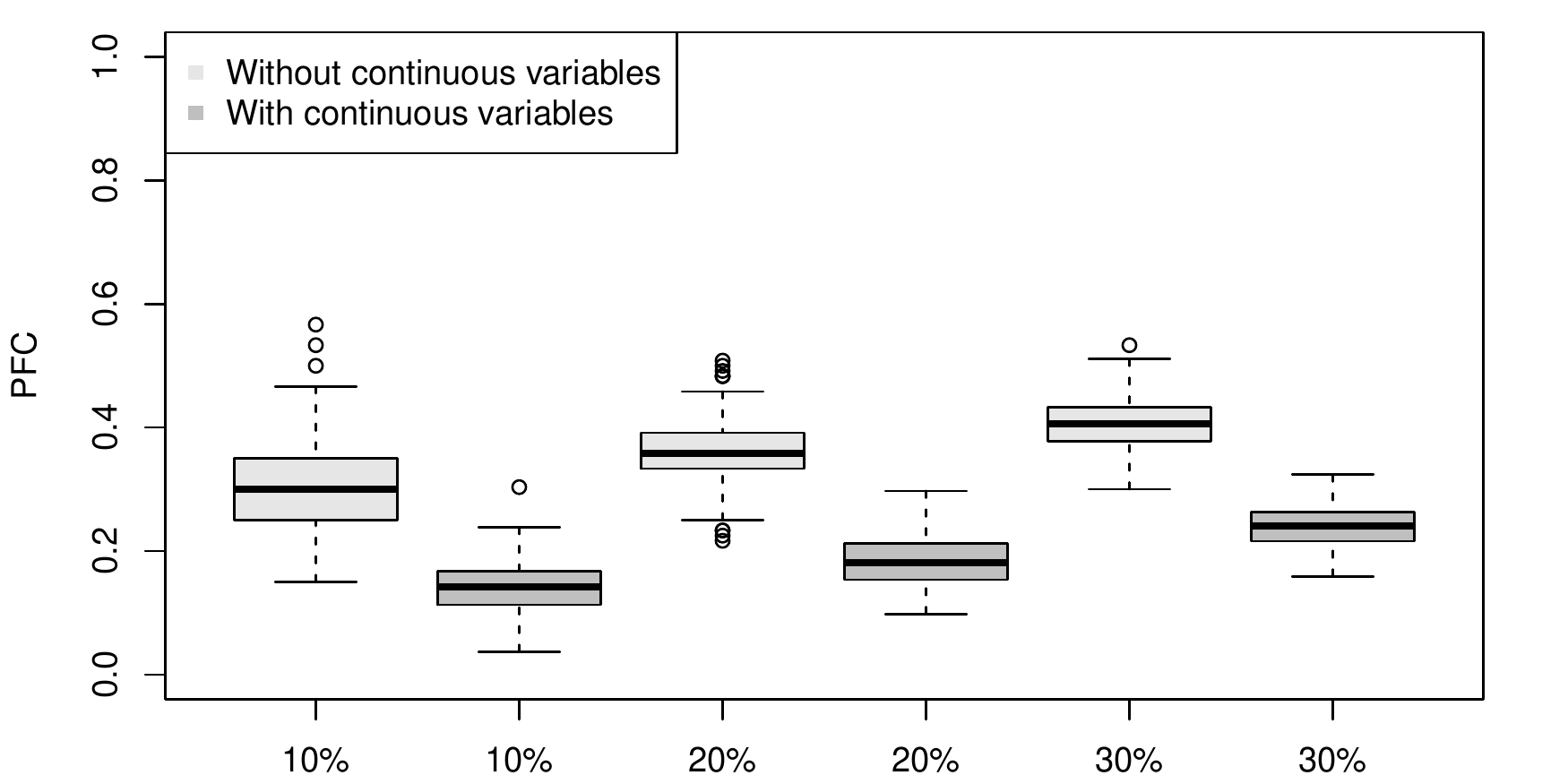}
  \caption{Distribution of the $PFC$ for different amounts of missing values (10\%, 20\%, 30\%). Light grey boxplots correspond to the error of imputation for categorical variables when only categorical variables are used, whereas grey boxplots correspond to the error when categorical and continuous variables are used.\label{main.prop.2}}
\end{figure}

Figure \ref{main.prop.2} compares the distributions of rates of misclassification according to the strategies (2) and (3). The results of the categorical variables are similar to those obtained for continuous variables: when the rate of missing data increases, the proportion of misclassification increases, but even for 30\% of missing data the imputation by FAMD gives better results than a random imputation (the error would be 0.66). Whatever the rate of missing data, taking into account the continuous variables for the imputation of categorical variables (boxplots light grey) reduce the proportion of misclassification.

\noindent \textit{Remark}: It is possible in theory to perfectly impute a categorical variable thanks to a continuous variable. On the contrary, it is difficult to impute the continuous variables with only categorical variables. For example, using $K_2$ categorical variables with $q$ categories each can produce only $q^{K_2}$ distinct imputations. Consequently, it cannot reflect the possible values that the continuous variable can take. However, in practice it is often a reasonable imputation.

\subsection{Influence of the relationships between variables}

\subsubsection{Linear and nonlinear relationships}
\begin{figure}[!hbt]
\begin{center}
\includegraphics[width=0.48\textwidth]{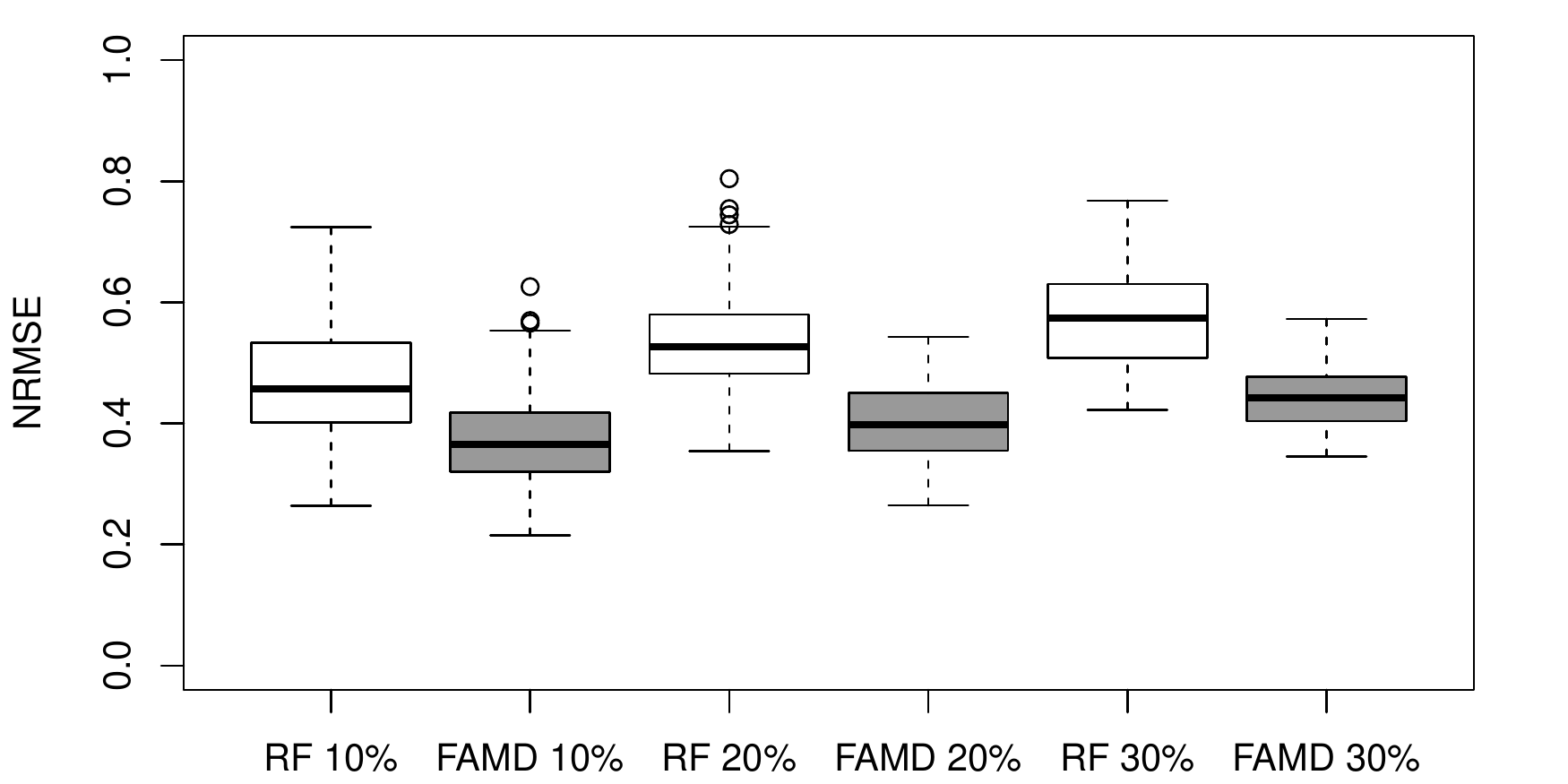}~~\includegraphics[width=0.48\textwidth]{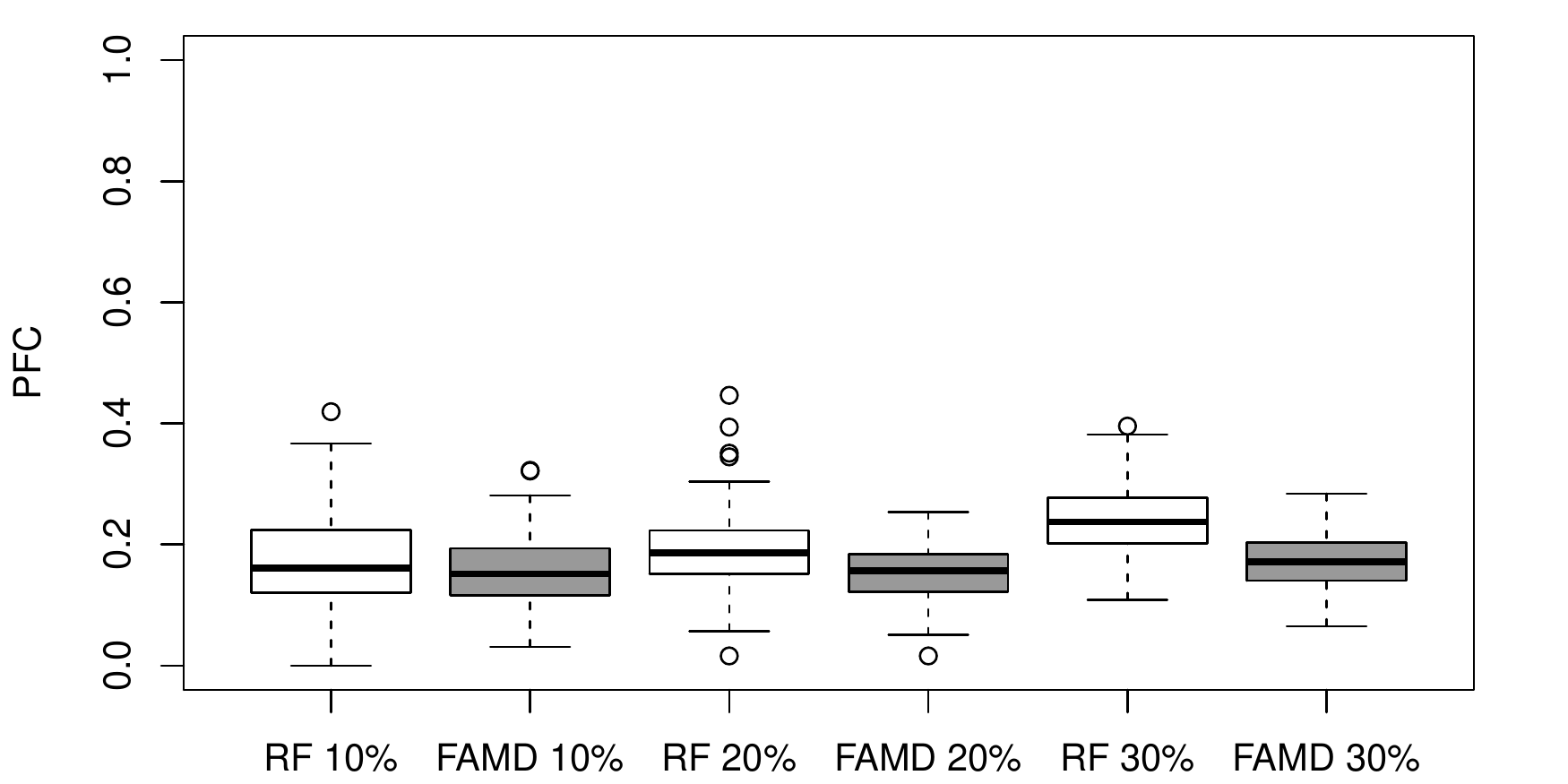}
\caption{Distribution of the NRMSE (left) and of the PFC (right) when the relationships between variables are linear for different amounts of missing values (10\%, 20\%, 30\%). White boxplots correspond to the imputation error for the algorithm based on random forests (RF) and grey boxplots to the imputation error for iterative FAMD.\label{simu_linear}}
\end{center}
\end{figure}

FAMD can be seen as a particular PCA and PCA is based on linear relationships between variables. When there are strong linear relationships between continuous variables, the imputation of these variables with iterative FAMD will thus be accurate. To illustrate this behaviour, a dataset is generated according to the simulation process with $S'=1$ initial variable from which 2 continuous variables and 3 categorical variables with 4 categories are built. 
The imputation by FAMD (Figure~\ref{simu_linear}, boxplots in grey) is compared to the imputation based on random forests (Figure~\ref{simu_linear}, boxplots in white). 
The error on continuous variables as well as the error on categorical variables with iterative FAMD is very small and is really smaller than the error of the algorithm based on random forests. Moreover, when the percentage of missing values increases, the error with the iterative FAMD algorithm increases slightly, whereas the error with the algorithm based on random forests increases more. Such results are representatives of all the results obtained with different datasets.

\begin{figure}[!hbt]
\begin{center}
\includegraphics[width=.49\textwidth]{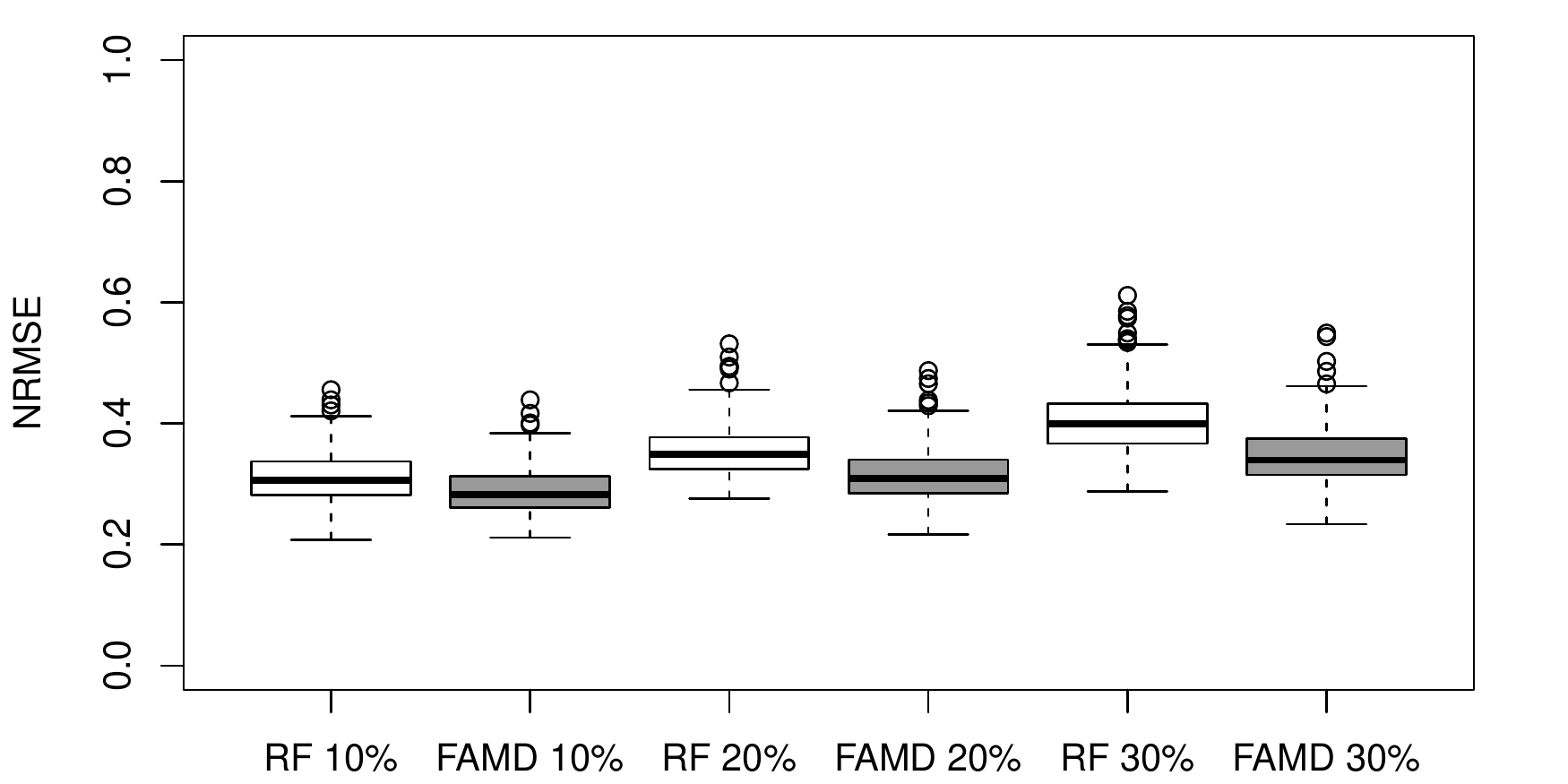}\hfill \includegraphics[width=.49\textwidth]{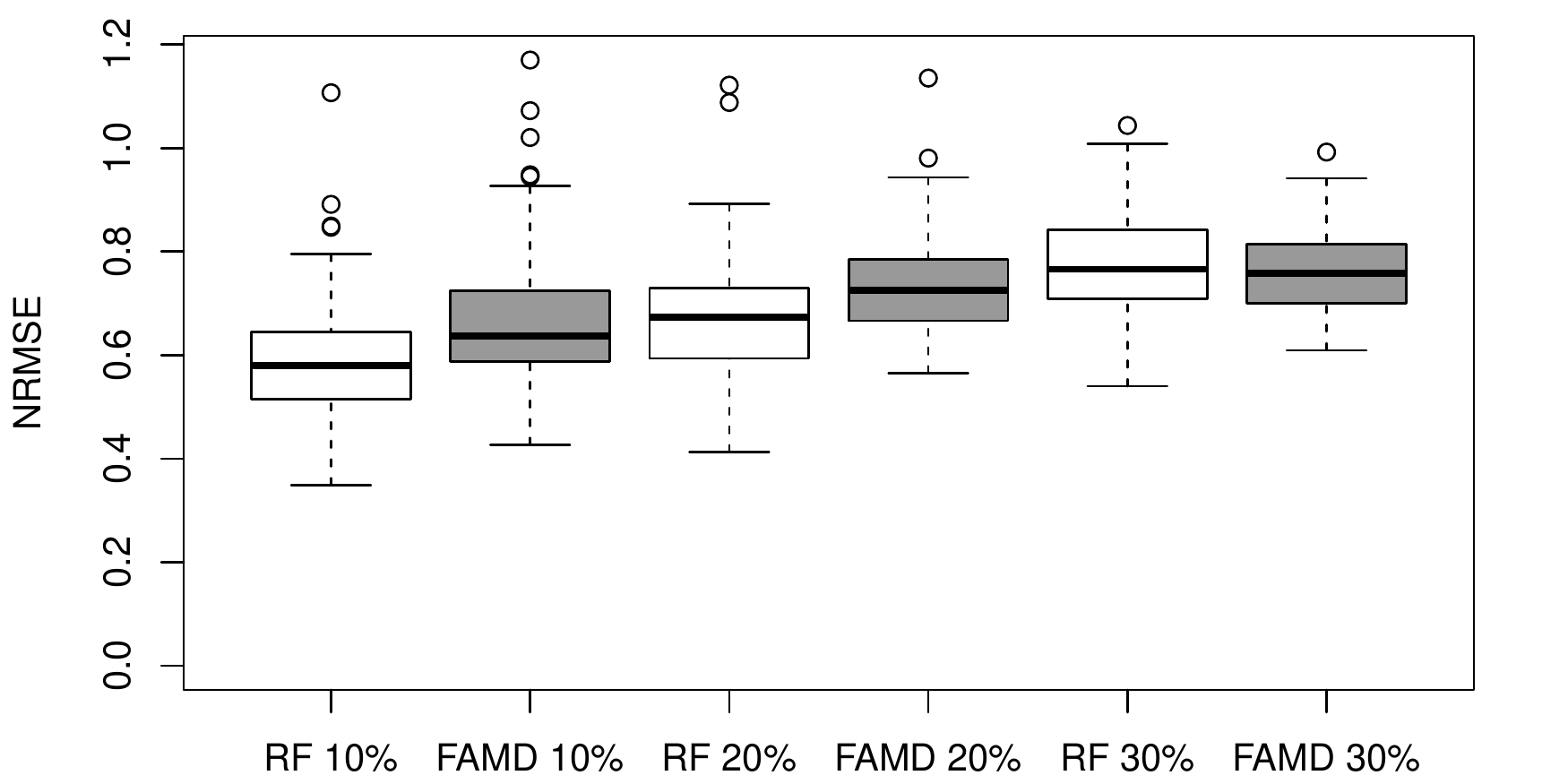}
\caption{Distribution of the NRMSE when the relationships between variables are linear (left) and nonlinear (right) for different amounts of missing values (10\%, 20\%, 30\%). White boxplots correspond to the imputation error for the algorithm based on random forests (RF), and grey boxplots to the imputation error for iterative FAMD.\label{relation}}
\end{center}
\end{figure}

It seems at first place more difficult to impute missing values with iterative FAMD when there are non linear relationships between continuous variables. However, due to the presence of categorical variables, these relationships can be taken into account. Indeed, the principal components of FAMD are linear combinations of the continuous variables and of the columns of the indicator matrix (equation \ref{crit.r2.eta2}). A linear combination of dummy variables may approximate a nonlinear function of a variable by a piecewise constant function.
To illustrate this behaviour, a dataset is generated with $S'=1$ from which 3 continuous variables and 1 categorical variable with 10 categories are built. First, we apply the methods on this data and the results illustrated Figure~\ref{relation} on the left are in accordance with the previous ones: the imputation with FAMD is very accurate when there are linear relationships. Then we take the same dataset but the second continuous variable is squared and the cosine function is applied to the third variable. The results obtained by iterative FAMD are in such a case worse than the ones obtained by the algorithm based on random forests (Figure~\ref{relation}, graph on the right), which is known to deal properly with nonlinear relationships. However, the differences remain reasonable.

Remark: if in practice, one suspects nonlinear relationships between continuous variables, a solution can be to create new categorical variables by cutting the continuous variables into categories.
\subsubsection{Taking into account interactions between categorical variables}
Other relationships between variables can be challenging. FAMD is based on relationships between pairs of variables. Consequently data including complex interactions could make the imputation difficult. To illustrate this behaviour, a specific dataset with 3 variables (1 continuous and 2 categorical) illustrated in Figure \ref{pattern} is constructed. It consists in a fractional factorial design $3^{3-1}$ that is replicated several times (stacked by columns below each other). Variables are pairwise independent but there are interactions between them. 
\begin{figure}[!hbt]
\small
\begin{center}
$\begin{array}{cccc}
   & \bfx_1   &\bfx_2 &\bfx_3\\
 1 &a  &a&  1 \\
 2 &b  &a&  2 \\
 3 &c  &a&  3 \\
 4 &a  &b& 2\\
 5 &b  &b& 3\\
 6 &c  &b& 1 \\
 7 &a  &c&  3 \\
 8 &b & c&  1 \\
 9&c&  c & 2 \\ \hdashline
  10 &a  &a&  1 \\
 11 &b  &a&  2 \\
 12 &c  &a& 3 \\
 13 &a  &b&  2\\
 14 &b  &b&  3\\
 15 &c  &b& 1 \\
 16 &a  &c&  3 \\
 17 &b & c&  1 \\
 18&c&  c &2 \\ \hdashline
 ...&&&
\end{array}$
\caption{Dataset with interaction generated with a fractional factorial design $3^{3-1}$; the defining relation of the fractional design is $I=123$.\label{pattern}}
\end{center}
\end{figure}
\begin{figure}[!hbt]
\begin{center}
\includegraphics[width=0.48\textwidth]{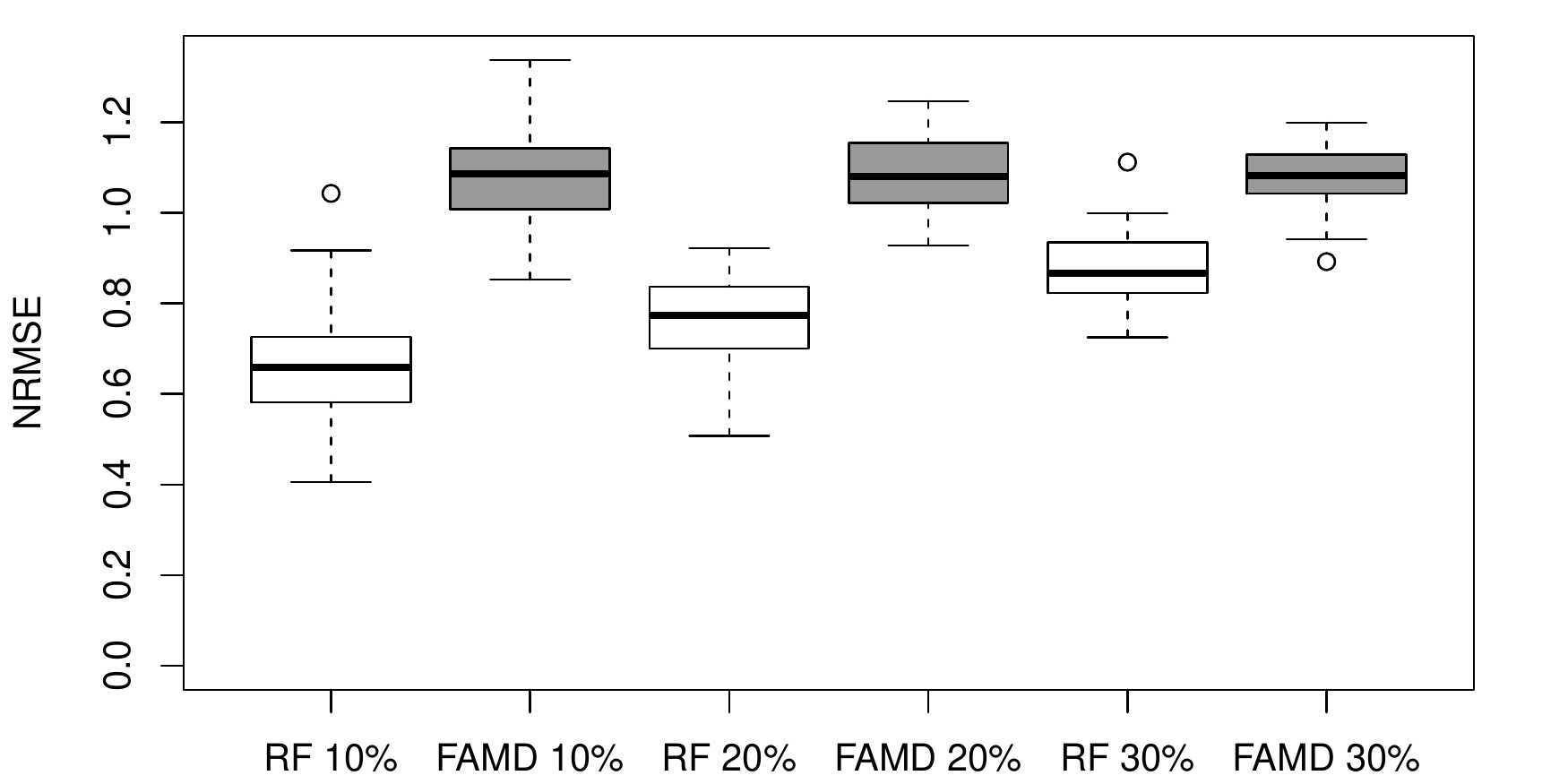}~~\includegraphics[width=0.48\textwidth]{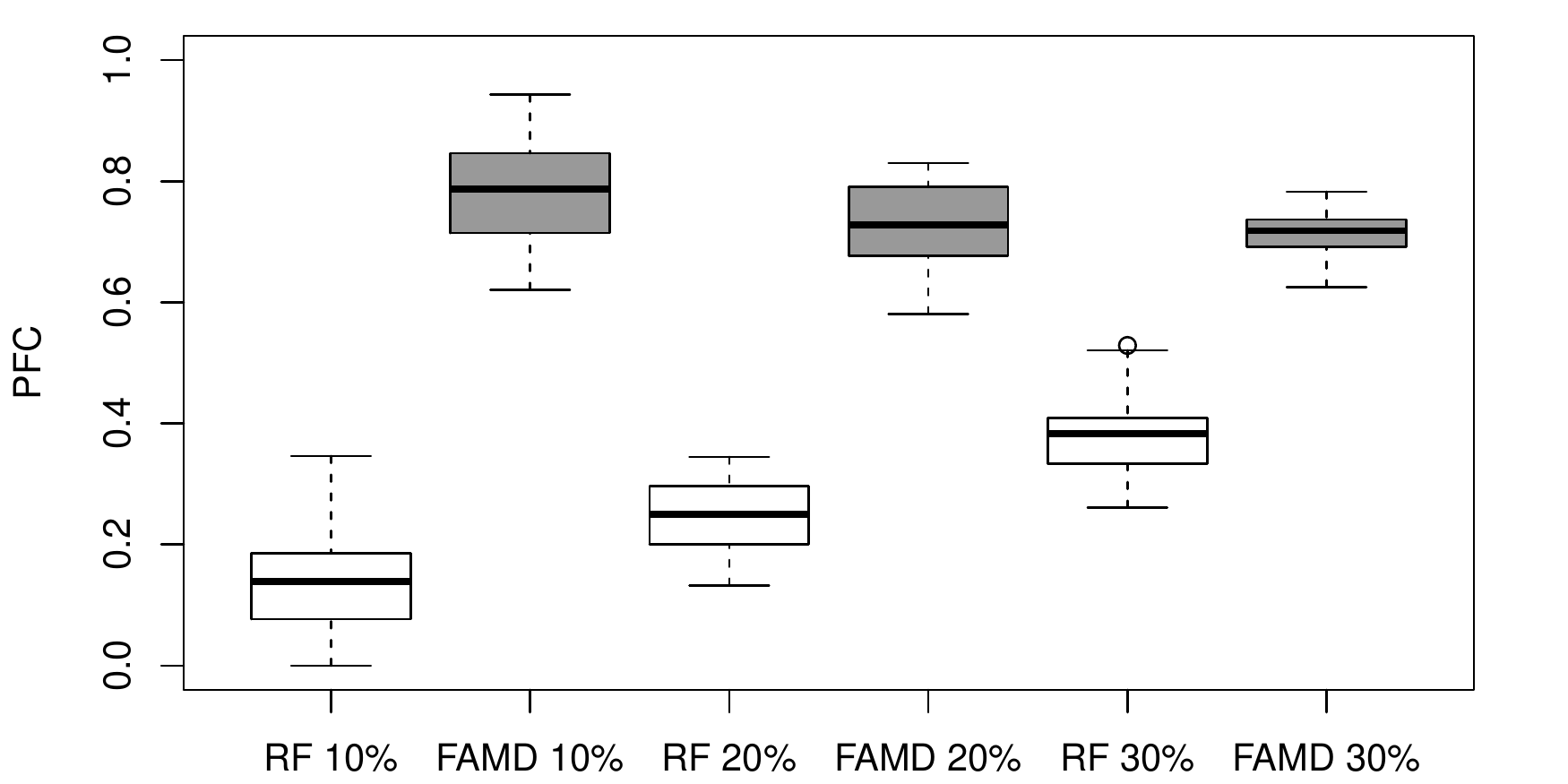}
\caption{Distribution of the NRMSE (left) and of the PFC (right) when there are interactions between variables. Results are given for different amounts of missing values (10\%, 20\%, 30\%). White boxplots correspond to the imputation error for the algorithm based on random forests (RF) and grey boxplots to the imputation error for iterative FAMD.\label{interaction}}
\end{center}
\end{figure}

The quality of the imputation of the continuous variable and of the categorical variables is poor with iterative FAMD (Figure~\ref{interaction}). It is closed to the mean imputation for the continuous variable and to the imputation by the proportion of the categories for the categorical variables.
The imputation based on random forests takes into account the interactions between variables and provides better results. It can be seen as a drawback of the imputation method based on FAMD.
However, we can address this problem by introducing an additional variable in the dataset that corresponds to the interaction, for example, by creating a variable $x_4$ which has 9 levels ``aa'', ``ba", ``ca", etc. It thus boils down to the case without interaction and the quality of the imputation will be very good.
\subsection{Imputation of rare categories}
A FAMD performed on a complete dataset weights each category by the inverse of the number of individuals taking this category (section~\ref{FAMD_complete_case}). Thus FAMD assigns a lot of variance to rare categories both in the cloud of categories as well as in the cloud of individuals in the initial space. Consequently, rare categories are privileged when constructing the dimensions of variability and because the algorithm uses the dimensions to impute the data, rare categories may be well predicted.

In order to illustrate the ability of the method to impute rare categories, simulations have been conducted with 2 continuous variables, 3 categorical variables and 3 categories per categorical variables. The frequencies of the categories are respectively equal to $(1/3,1/3,1/3)$ for one categorical variable and to $(f,(1-f)/2,(1-f)/2)$ for the 2 other categorical variables, with $f$ a frequency that varies between 0.004 and 0.1.
The rare categories (associated to the frequency $f$) of these 2 categorical variables are linked together in the sense that they are taken by the same individuals. Then a rare value is suppressed for one individual on one of the two categorical variables and the imputation algorithms are performed. It allows us only to focus on the prediction of a rare category.
Simulations are performed for different number of individuals and consequently, for a same frequency $f$, we expect that it would be easier to recover the true category when the number of individuals is large because it concerns more individuals.
The results (Table~\ref{rare}) show that the algorithm based on FAMD is good to recover the rare category. The advantage of this algorithm is all the more important as the category is rare compared to the algorithm based on random forests.

\begin{table}[!hbt]
\begin{center}
\begin{tabular}{cccc}
\hline
\multicolumn{1}{p{2cm}}{Number of individuals}& $f$&FAMD&Random forests\\
\hline
100  & 10\%   &0.060 &0.096 \\
100  & 4\%    &0.082 &0.173 \\

1000 & 10\%   &0.042 &0.041 \\
1000 & 4\%    &0.060 &0.071 \\
1000 & 1\%    &0.074 &0.167 \\
1000 & 0.4\% &0.107 &0.241 \\
\hline
\end{tabular}
\caption{Percentage of error (PFC) over 1000 simulations when recovering a rare category for datasets with different numbers of individuals, different frequencies for the rare category ($f$). Results are given for the imputation with FAMD and with the algorithm based on random forests.\label{rare}}
\end{center}
\end{table}

\subsection{Choice of the number of dimensions}
At each iteration of the iterative FAMD algorithm, data are reconstructed using only the $S$ first dimensions (step 2.b). If $S$ is too small then a part of relevant information is lost and cannot be used for the imputation. On the contrary, if $S$ is too large, then noise is considered as signal, which may lead to instability on the imputations. The number $S$ is thus an important parameter of the algorithm, which has to be chosen \textit{a priori}. In this section we focus in choosing this number from an incomplete mixed dataset.

First of all,  we can note that a categorical variable with $q_k$ categories evolves within a space with ($q_k-1$) dimensions. Therefore, it is impossible to predict the values of the categories with a choice of $S$ less than ($q_k-1$). This may be a clue that suggests choosing $S$.
However, it may happen that some of these ($q_k-1$) dimensions are not related to any other variables, especially when the number of categories is high. In this case, even if $S$ is large, it is impossible to impute correctly this variable and choosing such a high number of dimensions may lead to instability.

Many strategies are available in the literature to select a number of dimensions from a complete dataset in PCA. Cross-validation \citep{Bro08} or approximation of cross-validation such as generalized cross-validation \citep{Jossenbaxe11} are methods showing good performances. These methods have the advantage that they can be directly applied on incomplete data. Since their extension is straightforward for FAMD, we use, in practice, cross-validation to select the number of dimensions. However, this topic may deserve more research. Consequently, it is important to assess the impact of a wrong choice for the number of dimensions on the results.

We consider a dataset generated from $S'=2$ groups of orthogonal variables ($K^{1'}=8$ variables, 4 continuous and 4 categorical variables, and $K^{2'}=4$ variables, 2 continuous and 2 categorical variables). Each categorical variable has 3 categories. Consequently the underlying number of dimensions of this dataset is $S=4$.
\begin{figure}[!hbtp]
\begin{center}
\includegraphics[width=0.48\textwidth,trim = 0mm 5mm 0mm 0mm, clip]{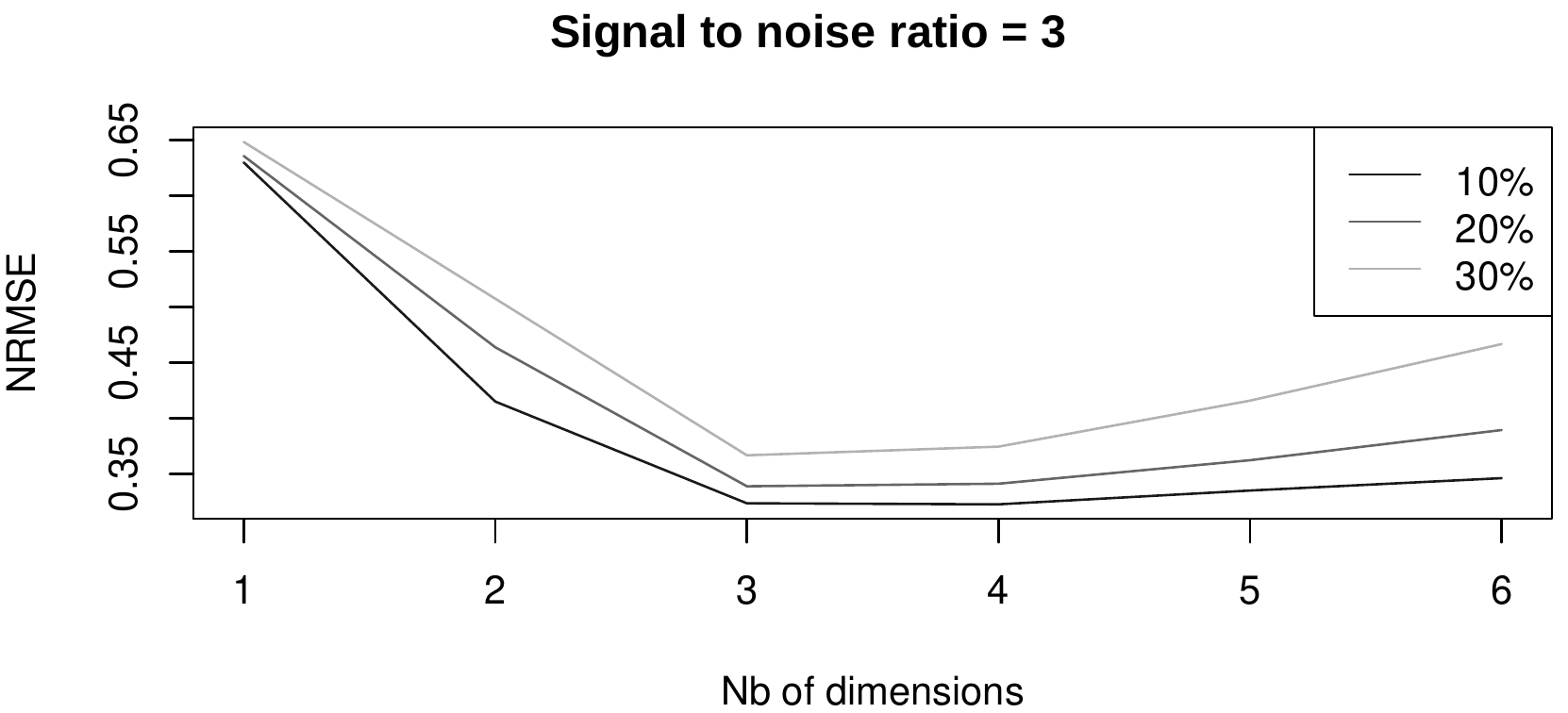}\hfill\includegraphics[width=0.48\textwidth,trim = 0mm 5mm 0mm 0mm, clip]{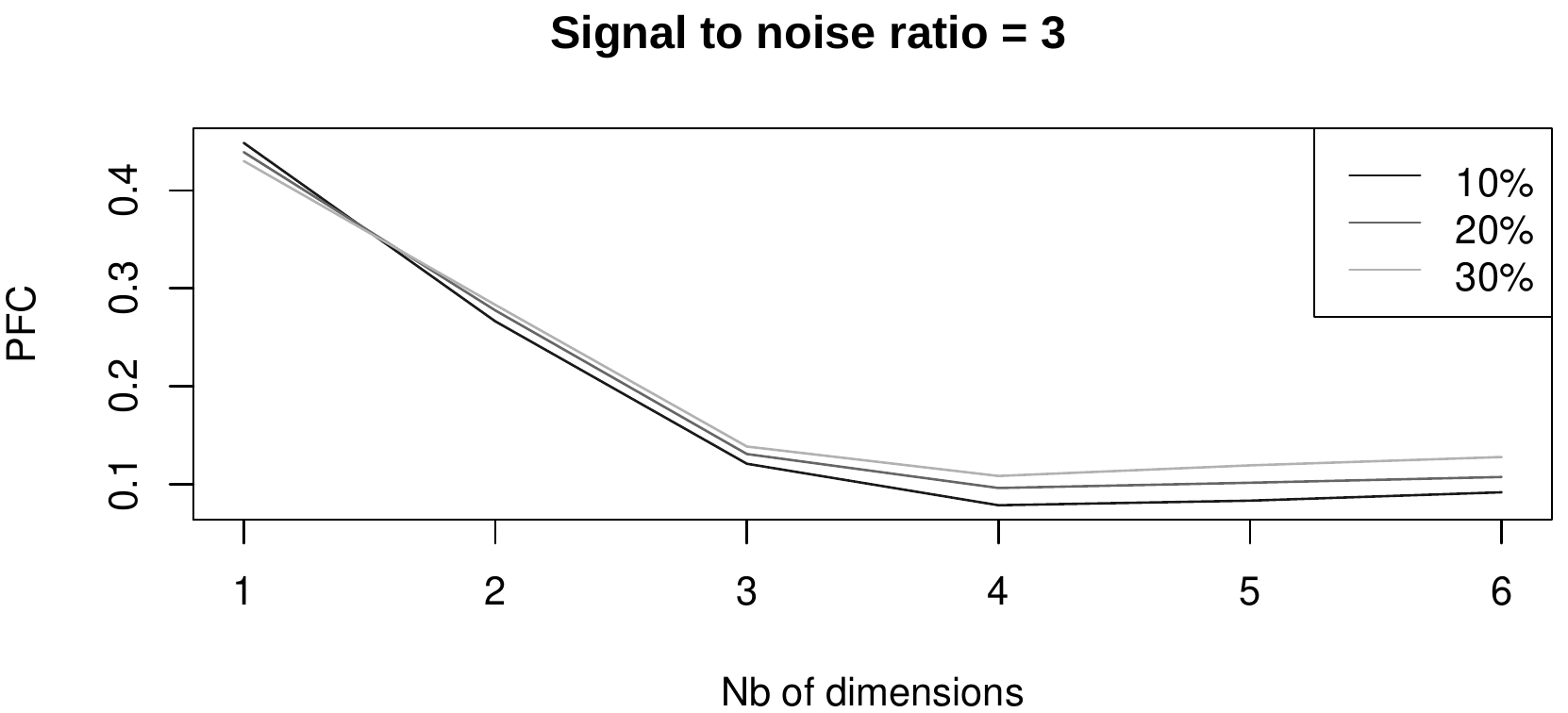}
\includegraphics[width=0.48\textwidth,trim = 0mm 5mm 0mm 0mm, clip]{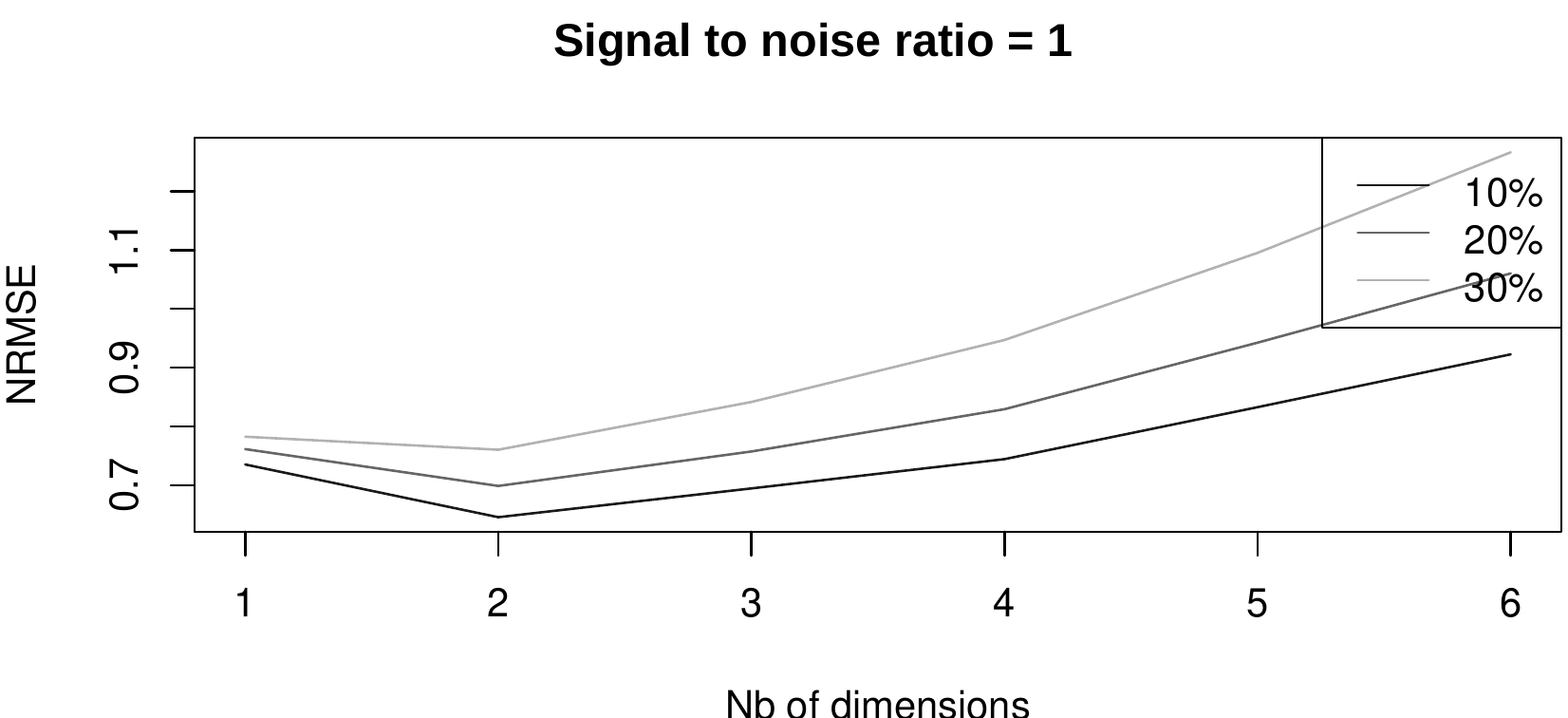}\hfill\includegraphics[width=0.48\textwidth,trim = 0mm 5mm 0mm 0mm, clip]{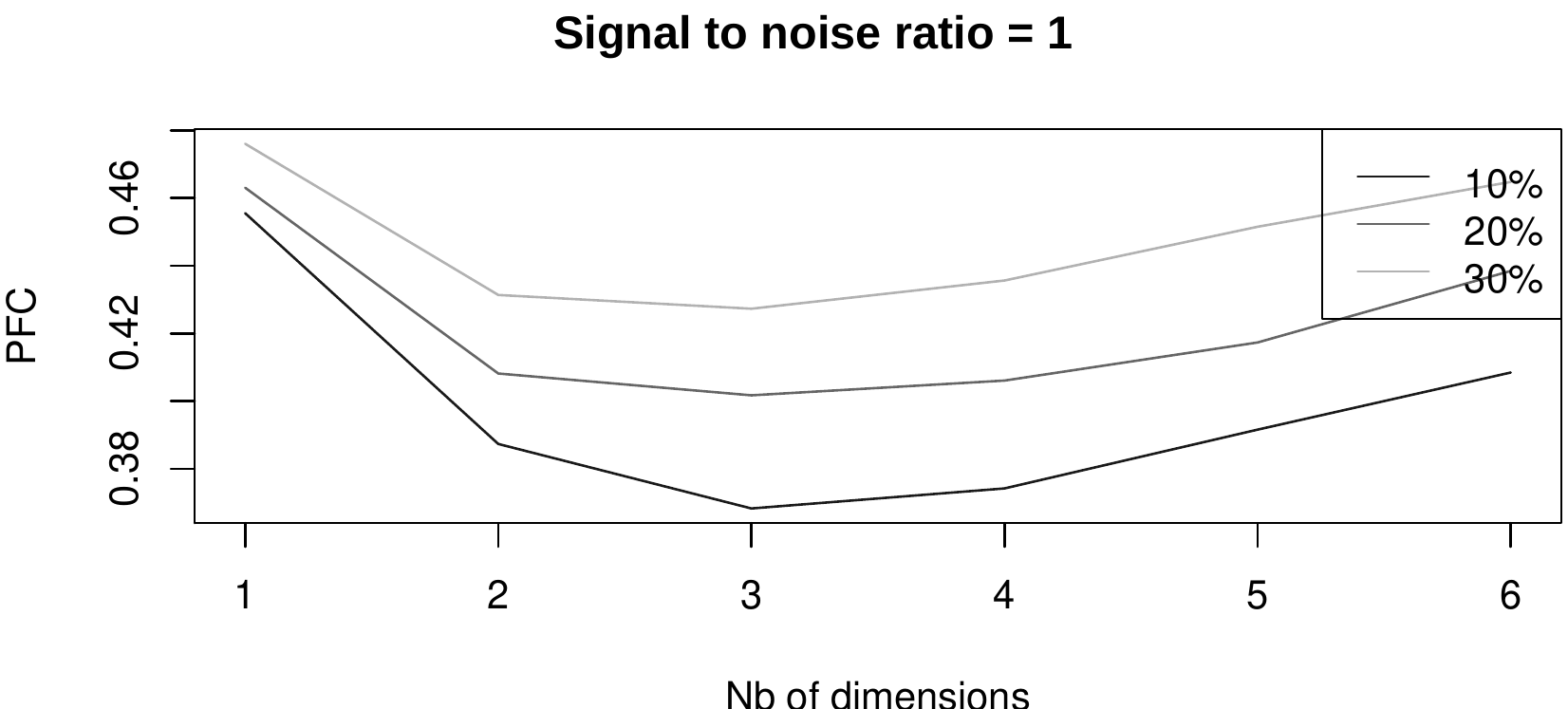}
\caption{Average error of imputation over 200 simulations according to the number of dimensions used in the algorithm and for 3 amounts of missing values (10\%, 20\%, 30\%): error for the continuous variables on the left and for the categorical variables on the right. The signal to noise ratio equals 3 for the simulations represented on the top, and 1 for the simulations represented on the bottom.\label{nbaxe}}
\end{center}
\end{figure}
Figure \ref{nbaxe} shows the evolution of the average of the errors over 200 simulations according to $S$ for the continuous variables (graphs on the left) and for the categorical variables (graphs on the right). The graphs on the top are obtained for a structured dataset (SNR=3) whereas data are noisy (SNR=1) for the graphs on the bottom.
When the signal to noise ratio is high, the error for the categorical variables decreases until it reaches the optimal value 4 and then it increases slowly, whatever the percentage of missing values. The same comment can be made for the continuous variables even if the minimum is reached for $S=3$, since the continuous variables are only linked to the first 3 dimensions. The behaviour is very different when the signal to noise ratio is small. Indeed, even for the categorical variables, it is preferable to choose a number of dimensions that is less than the true number. This is all the more true as the percentage of missing values is important. This behaviour can be explained, since selecting a smaller number of dimensions can be regarded as performing stronger regularization which is a good strategy when the data are very noisy. Moreover taking a smaller number of dimension is acceptable because the true underlying dimensions can be considered as lost.
%Consequently, it is not possible to say that it is preferable to select more or less components than the true number, since the impact depends on the structure of the data. A good strategy is thus to resort to cross-validation techniques. 

\section{Comparison on real datasets}

The imputation method is evaluated on real datasets that cover many situations. They differ in terms of number of individuals, number of variables, number of categories for the categorical variables, and that come from different areas of application. Missing values are added at random from these complete sets and then the imputation is performed with iterative FAMD and with \citet{Buhlmann11} algorithm. Each configuration is simulated 200 times for three different percentages of missing data. The number of dimensions for the reconstruction step of the iterative FAMD algorithm is determined by cross-validation. This number is kept as fixed for the 200 simulations in order to save computational time. The evaluation is based on the following mixed datasets.
\paragraph{Tips}
This dataset, from the package \texttt{rggobi} \citep{rggobi} of the \texttt{R} software \citep{Rsoft}, concerns the tips given to a waiter in a restaurant in the U.S. in the early 1990s. The $K=8$ variables of the dataset focus on the price of the meal for $I=244$ customers, on the tip amount and the conditions under which the meal is taken (number of guests, time of day, etc.). There are $K_1 =3$ continuous variables and  $K_2=5$ categorical variables with between 2 and 6 levels.
\paragraph{BMI}
This dataset \citep{Lafaye11} studied body mass index of $I=152$ French children aged 3 to 4 years. The $K=6$ variables concern their morphology and the characteristics of their kindergarten ($K_1=4 $, $K_2=2$). All the categorical variables have two levels.
\paragraph{Ozone}
This dataset \citep{Cornillonenglish12} contains $I=112$ daily measurements of meteorological variables (wind speed, temperature, rainfall, etc.) and ozone concentration recorded in Rennes (France) during the summer 2001. There are $K_1 =11$ continuous variables and $K_2 =2$ categorical variables with 2 or 4 levels.
\paragraph{German Breast Cancer Study Group (GBSG)}
This dataset, from the package \texttt{ipred} \citep{ipred} of the \texttt{R} software, described $I = 686$ women with breast cancer through variables including on the status of the tumours and the hormonal system of the patient ($K_1=7$, $K_2=3$). Categorical variables have between 2 and 3 levels.\\ \\
\begin{figure}[!hbtp]
\begin{tabular}{m{0.8cm}m{4.8cm}m{4.8cm}}
{\small Tips}&\includegraphics[width=0.45\textwidth]{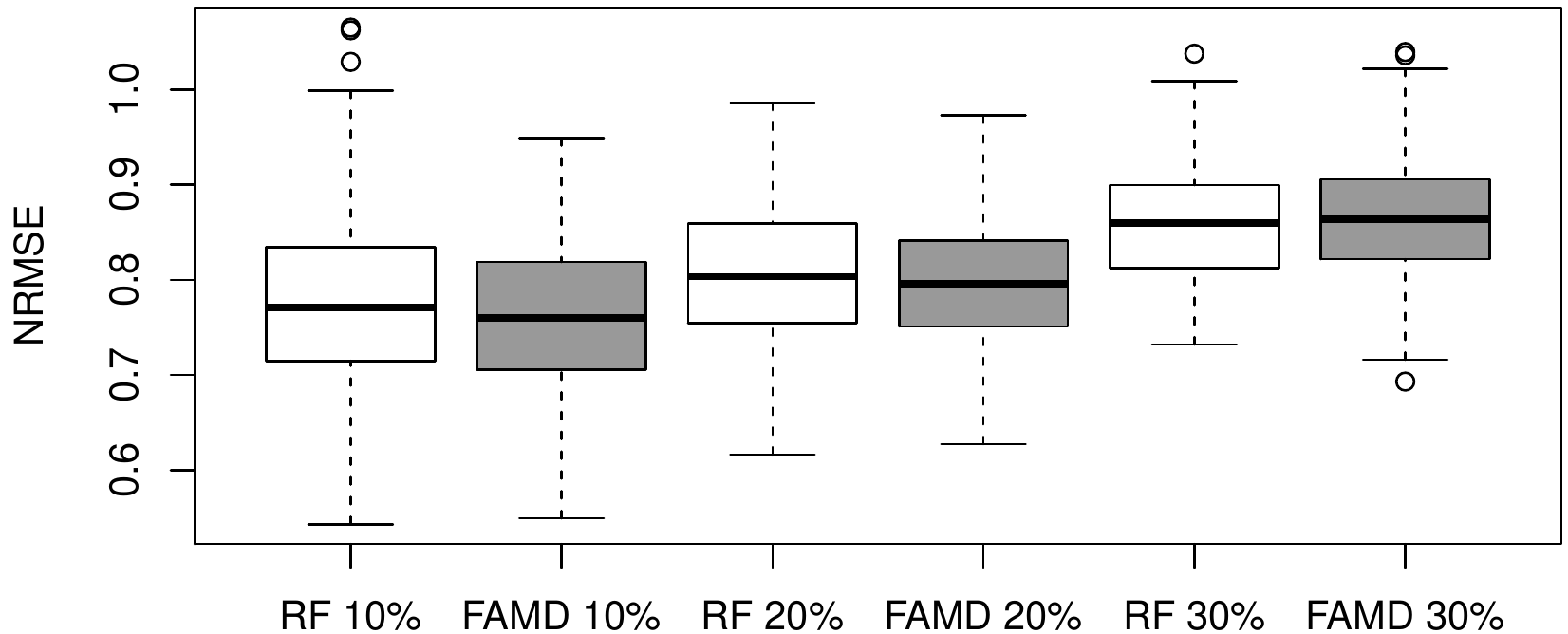}& \includegraphics[width=0.45\textwidth]{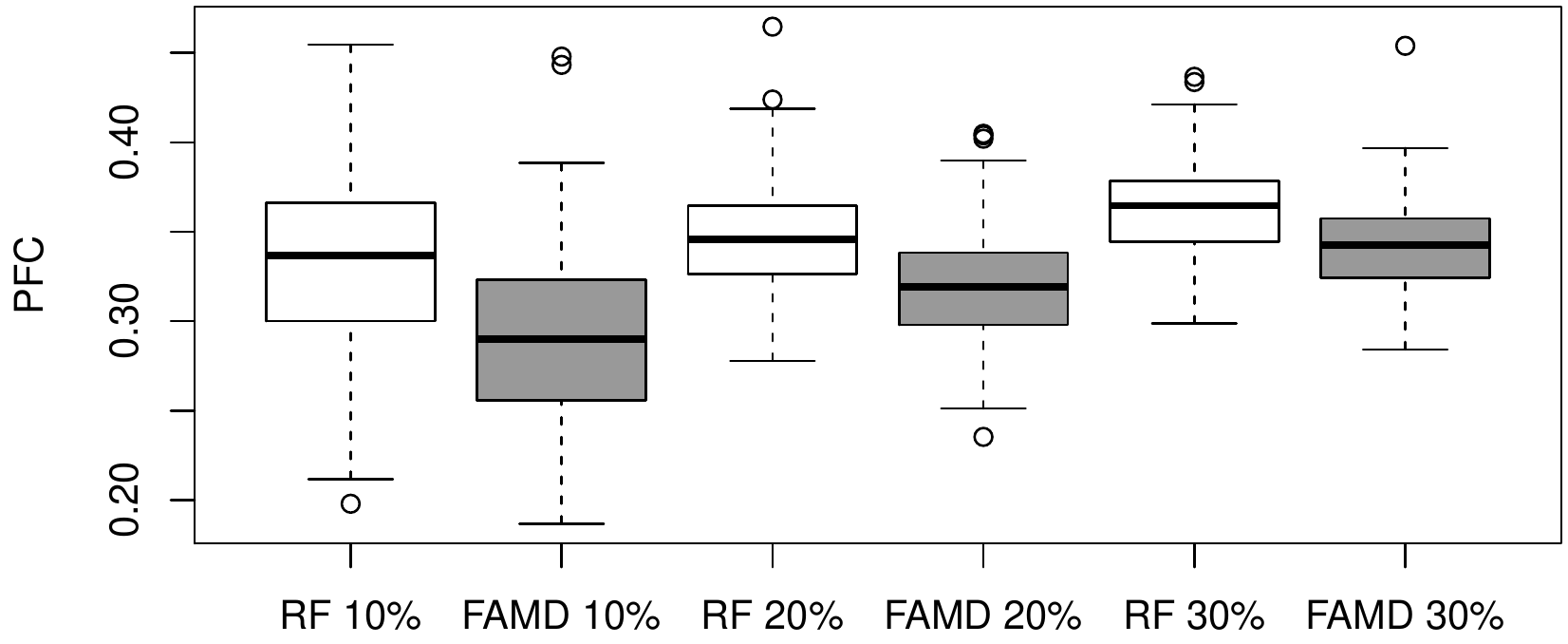}\\
{\small BMI}&\includegraphics[width=0.45\textwidth]{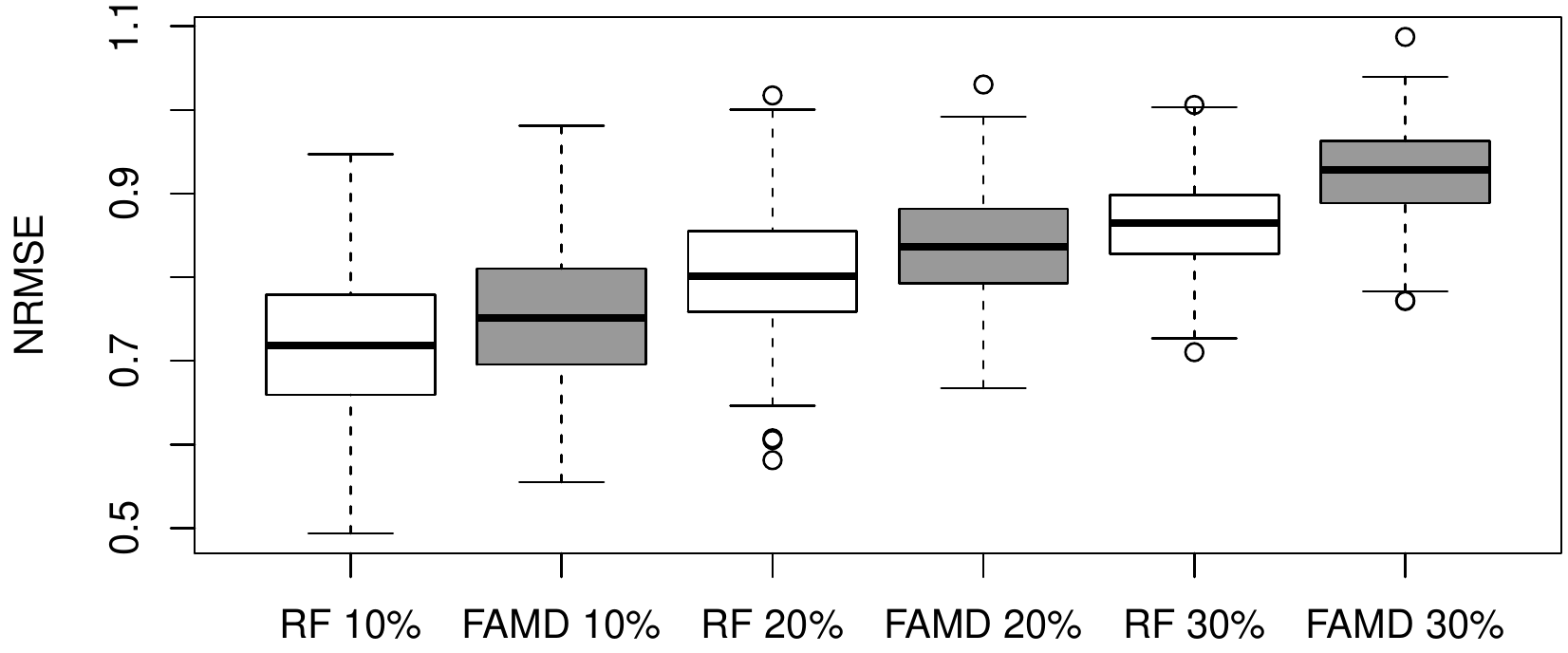}& \includegraphics[width=0.45\textwidth]{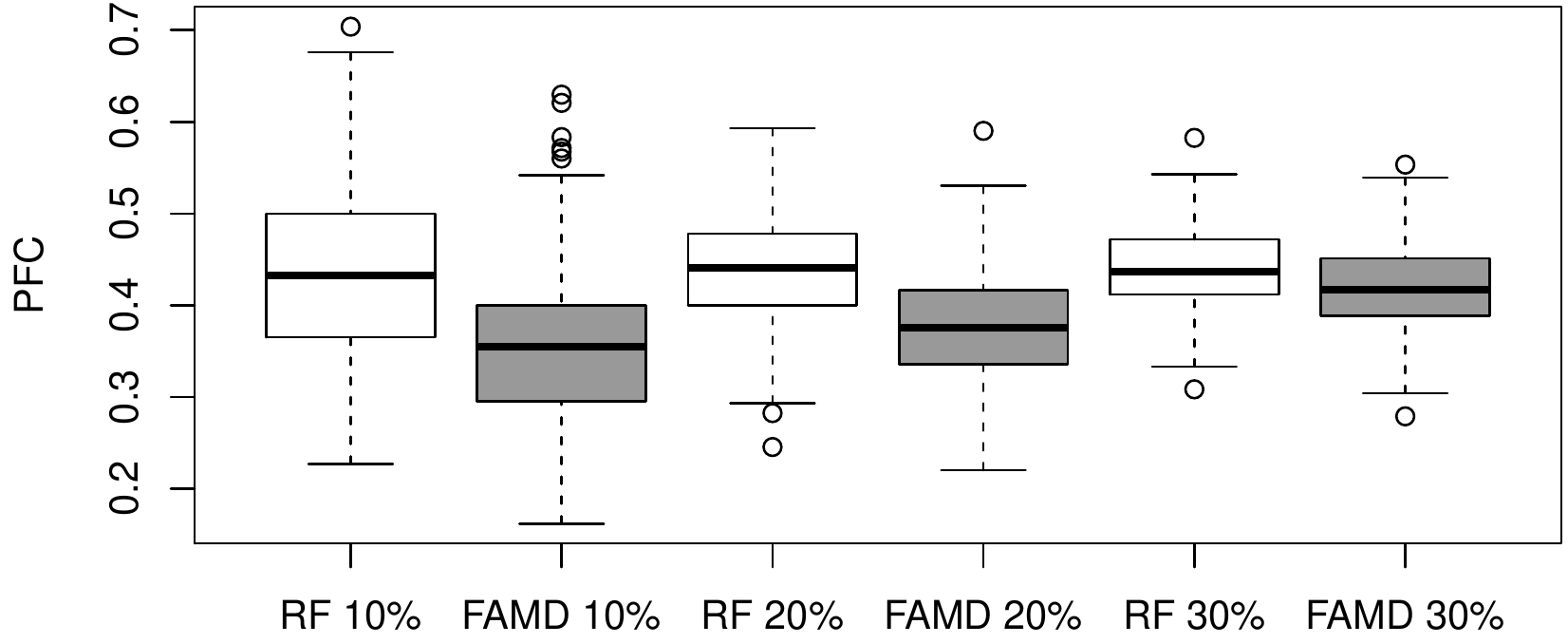}\\
{\small GBSG}&\includegraphics[width=0.45\textwidth]{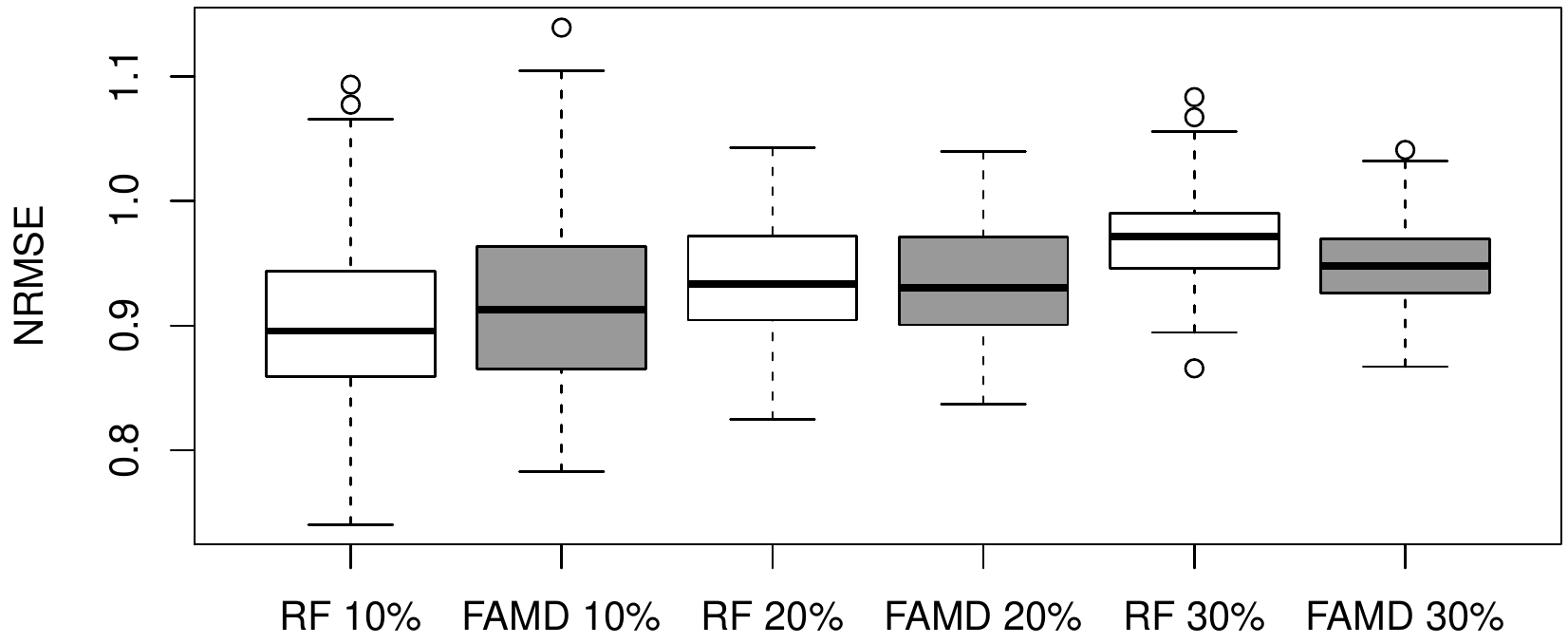}& \includegraphics[width=0.45\textwidth]{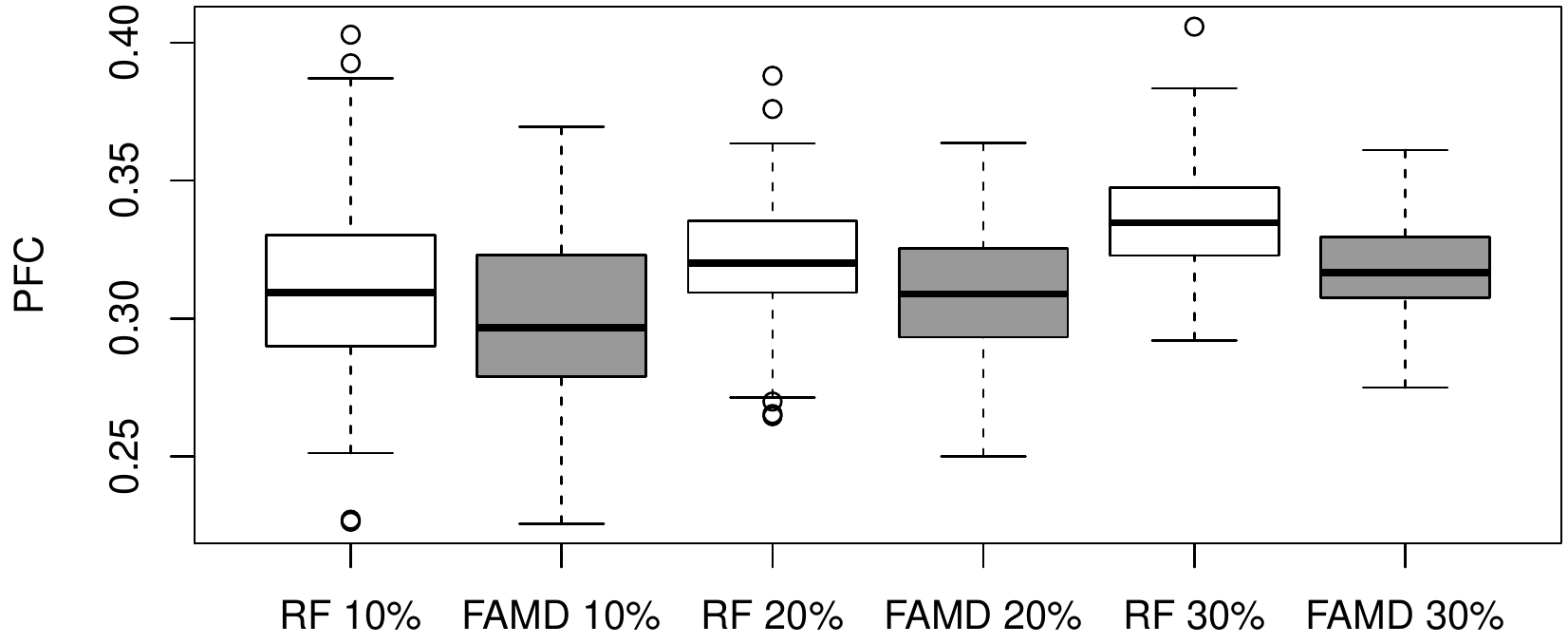}\\
{\small Ozone}&\includegraphics[width=0.45\textwidth]{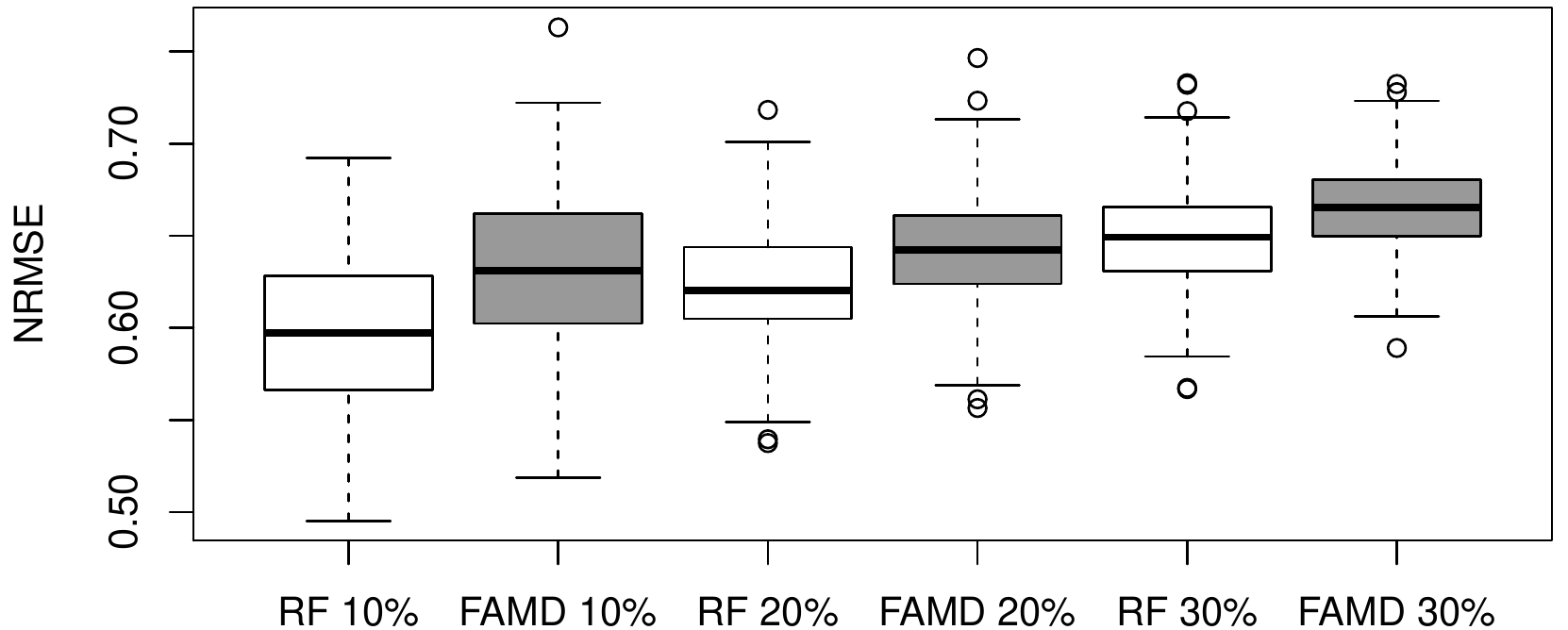}& \includegraphics[width=0.45\textwidth]{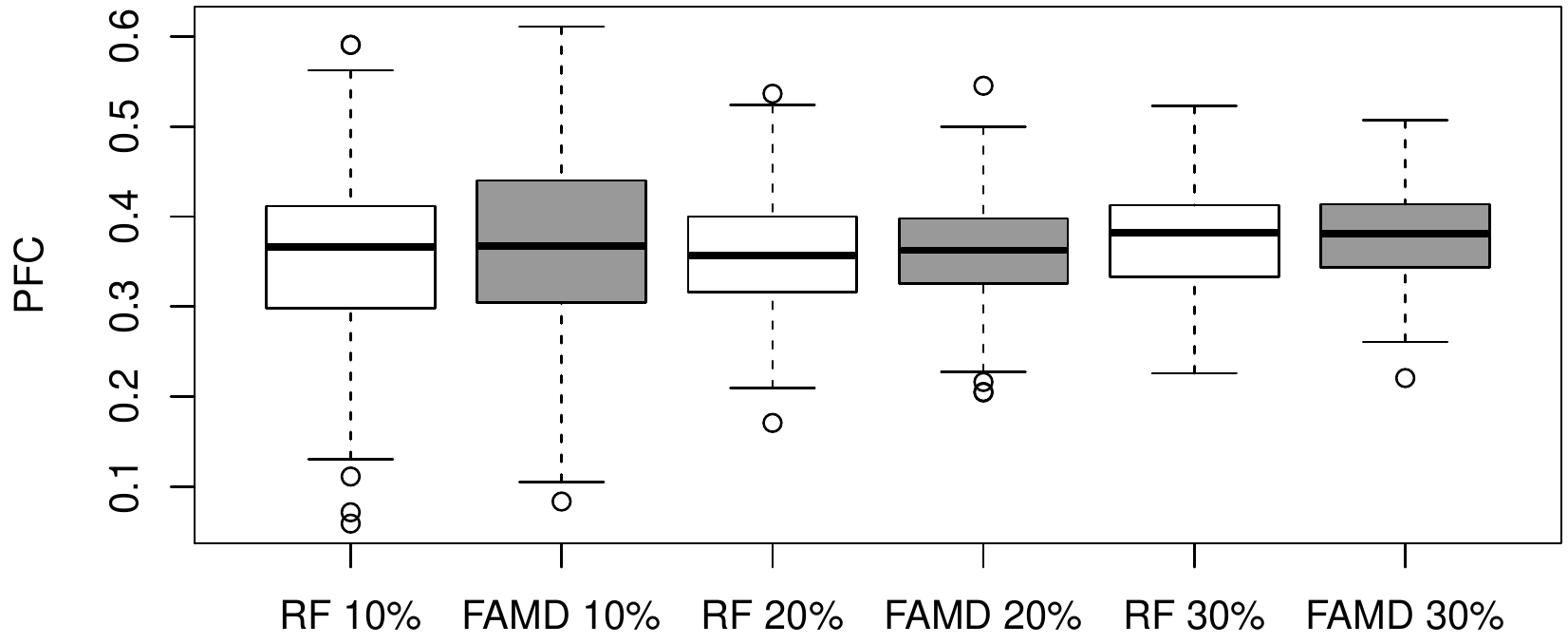}\\
\end{tabular}
\caption{Distribution of the NRMSE (left) and of the PFC (right) for different amounts of missing values (10\%, 20\%, 30\%) and for different datasets (Tips, BMI, GBSG, Ozone). White boxplots correspond to the imputation error of the algorithm based on random forests (RF) and grey boxplots correspond to the imputation error of iterative FAMD.\label{reels.mixte}}
\end{figure}
Imputation results for all the datasets are presented in Figure~\ref{reels.mixte}.
In general, the iterative FAMD provides a slightly better imputation than the one obtained by the algorithm based on random forests. The imputation with the iterative FAMD is more efficient on categorical variables. For the datasets on tips and BMI, the difference between the two methods is 5\%. For continuous variables, the advantage is more often for the algorithm based on random forests. On the datasets ozone and BMI, the difference between the errors reached 5\%.\\

These conclusions extend to the case of non mixed datasets. We now consider two of them: one continuous (Parkinson), the other categorical (Credit).
\paragraph{Parkinson}

This dataset \citep{Buhlmann11} contains $K=22$ measurements on the voice of $I =195$ patients having or not the Parkinson's disease. The response categorical variable sick/healthy is excluded for these simulations.
\paragraph{Credit}

This dataset \citep{Cornillonenglish12} focuses on $I=66$ customers profiles having subscribed to a consumer credit in a bank. The $K=11$ variables include the financial conditions under which the customer subscribes the credit as well as some socio-demographic characteristics. The number of levels for these variables is between 2 and 5. \\

\begin{figure}[!hbt]
\begin{center}
\includegraphics[width=.49\textwidth]{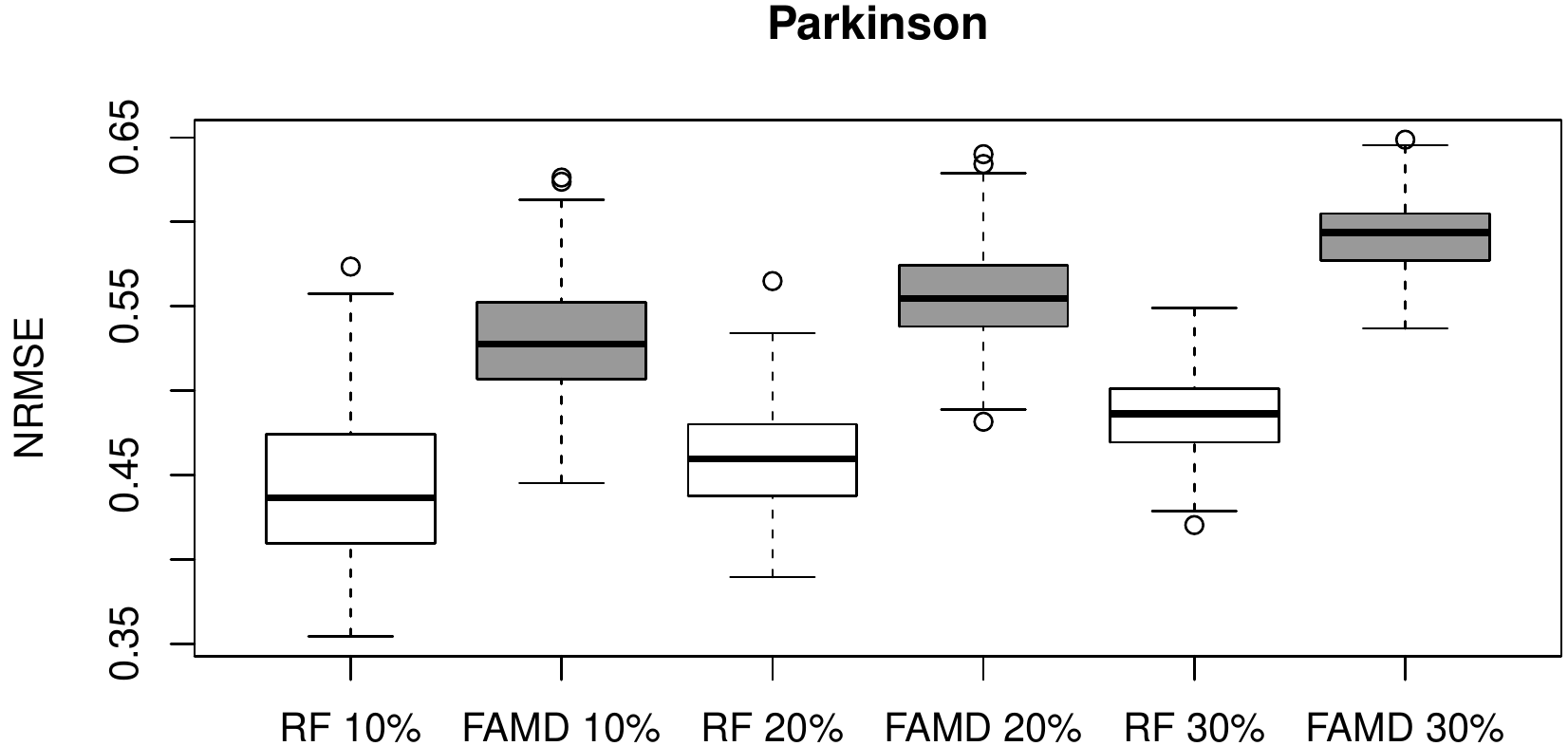} \includegraphics[width=.49\textwidth]{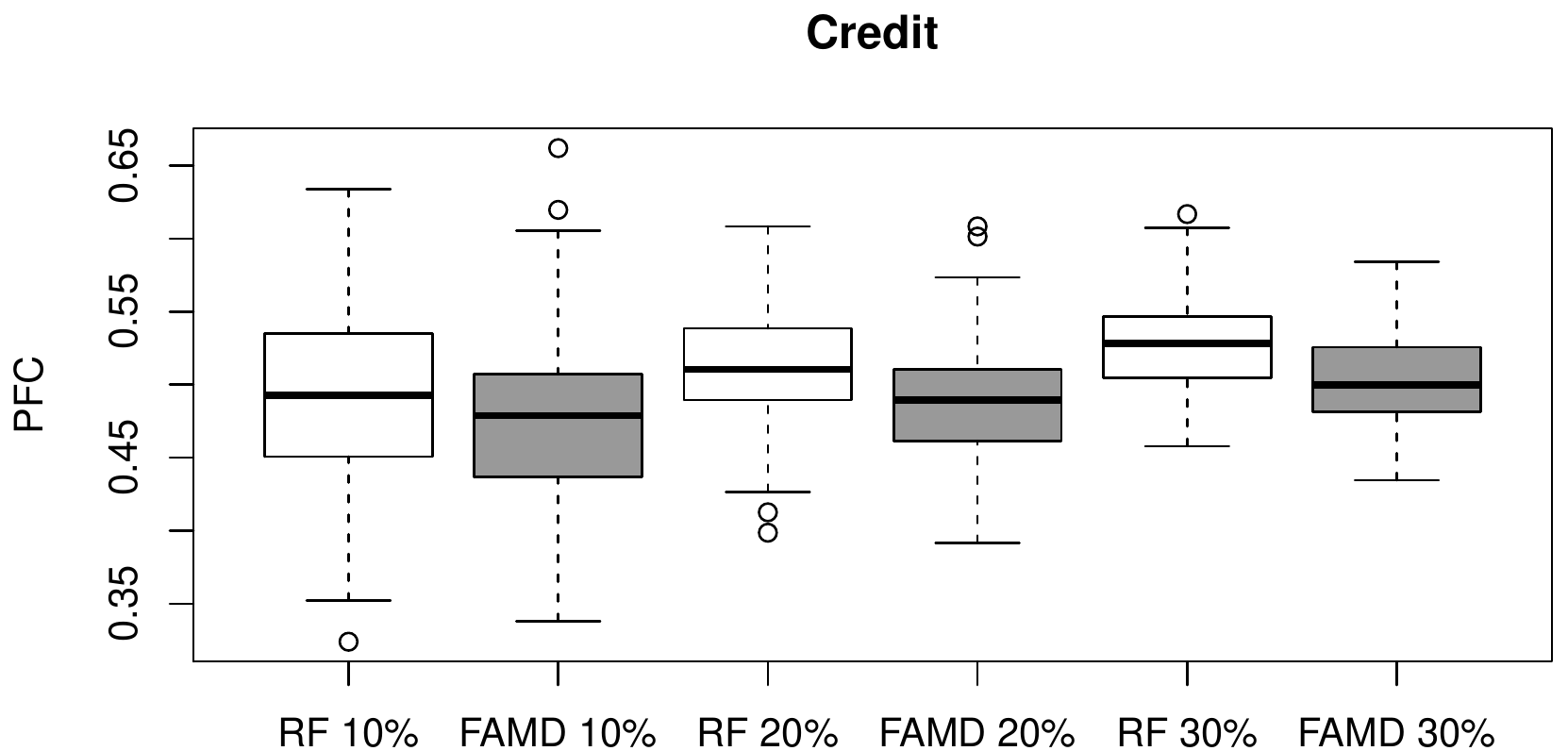}
\caption{Distribution of the error for 10\%, 20\%, 30\% of missing values for the datasets Parkinson (left) and Credit (right). White boxplots correspond to the imputation error of the algorithm based on random forests (RF) and grey boxplots correspond to the error of iterative FAMD.\label{reels.nonmixte}}
\end{center}
\end{figure}
\noindent The imputation of the continuous dataset (Parkinson) is much to the advantage of the random forest imputation with a gap of about 10 \% between errors (Figure~\ref{reels.nonmixte} on the left). We thus find the same results about continuous variables as those of mixed-type data. This is accentuated by the fact that relationships between variables are here non-linear. The lack of categorical variables implies that it is not possible to impute these variables correctly with iterative FAMD.
For the categorical dataset (Credit), the errors (Figure~\ref{reels.nonmixte} on the right) are smaller with the iterative FAMD algorithm, whatever the percentage of missing values. These results are also similar to those observed in the mixed-type data.
\section*{Conclusion}

The new imputation method proposed in this paper is based on the principal components method factorial analysis for mixed data and allows one to simultaneously impute missing data taking into account the similarities between individuals and the links between continuous and categorical variables.
It gives particularly good predictions for the missing entries for the categorical variables and especially for rare categories and when there are linear relationships between variables. In addition, the method provides good results compared to the method based on random forests both in terms of quality imputation and computational time.

The iterative FAMD algorithm requires a tuning parameter which is the number of dimensions used to reconstruct the data. In practice, cross-validation can be used to select this number even if this strategy is time-consuming. Approximation of cross-validation such as generalised cross-validation could be proposed to select this number without resorting to an intensive computational method.

The imputation method based on FAMD is implemented in the package \texttt{missMDA} \citep{MissMDA} of the \texttt{R} software. The function \texttt{imputeFAMD} takes as input the incomplete dataset and the number of dimensions used to reconstruct the data at each step of the algorithm. The function returns a table with the imputed mixed data as well as the table concatenating the imputed continuous variables and the imputed indicator matrix.

As with all methods of imputation, imputation quality deteriorates with increasing percentage of missing data. However, this deterioration depends on the structure of the dataset. Indeed, if the variables are not linked, a single missing value is already problematic. On the contrary, if the variables are strongly linked, very little data per individual are sufficient to impute the dataset. For this reason, it is of course not possible to offer a percentage of missing data below which imputation is acceptable and above which the imputation is no longer satisfactory. It would be therefore desirable to provide confidence intervals around the imputed values. We expect small confidence intervals when the data are very related which will lead to trust in the imputed values.

The proposed method is a method of single imputation. Like any single imputation method, it suffers from not taking into account the uncertainty associated with the prediction of missing values based upon observed values. Thus, if we apply a statistical method on the completed data table, the variability of the estimators will be underestimated. To avoid this problem, a solution is to use multiple imputation \citep{Little02}.
In the latter, different values are predicted for each missing value, which leads to several imputed datasets and the variability among the imputation reflects the variance of the prediction of each missing entry. The second step of multiple imputation consists in performing the statistical analysis on each completed dataset and its third step combines the results to obtain the estimators of the parameters and of their variability taking into account missing data uncertainty.

The proposed iterative FAMD imputation algorithm could be a first step in a multiple imputation method for mixed data. Such a method on the one hand would give confidence intervals for each imputed values and on the other hand would allow one to perform statistical analyses from incomplete mixed datasets.

\bibliographystyle{spbasic}
\bibliography{audigier}
\end{document}